\documentclass[pre,amsmath,amssymb,reprint, superscriptaddress,longbibliography]{revtex4-1}
\usepackage[utf8]{inputenc}
\usepackage{amsmath}
\usepackage{amssymb}
\usepackage{nicefrac}
\usepackage{relsize}
\usepackage{etoolbox}
\usepackage[shortlabels]{enumitem}
\usepackage{breqn} 
\usepackage{graphicx}
\usepackage{subfigure}
\usepackage{dcolumn}
\usepackage{bm}
\usepackage{float}
\usepackage{hyperref}

\makeatletter
\patchcmd\eq@setnumber{\stepcounter}{\refstepcounter}{}{%
  \errmessage{Patching \noexpand\eq@setnumber failed}%
}
\makeatother
\makeatletter
\let\cat@comma@active\@empty
\makeatother
\newcommand{\tlambda}{\widetilde{\lambda}}
\newcommand{\tA}{\widetilde{A}}
\newcommand{\pcrit}{p_c}
\newcommand{\lcrit}{\lambda_c}
\newcommand{\dinf}{d_{\infty}}
\newcommand{\scO}{\mathcal{O}}

\begin{document}

\title{Critical Network Cascades with Re-excitable nodes: Why tree-like approximations usually work, when they breakdown, and how to correct them}
\author{Sarthak Chandra}
\email{sarthakc@umd.edu}
\affiliation{Department of Physics, University of Maryland, College Park, MD, U.S.A.}
\affiliation{Institute for Research in Electronics and Applied Physics, University of Maryland, College Park, MD, U.S.A.}
\author{Edward Ott}
\affiliation{Department of Physics, University of Maryland, College Park, MD, U.S.A.}
\affiliation{Institute for Research in Electronics and Applied Physics, University of Maryland, College Park, MD, U.S.A.}
\affiliation{Department of Electrical and Computer Engineering, University of Maryland, College Park, MD, U.S.A.}
\author{Michelle Girvan}
\affiliation{Department of Physics, University of Maryland, College Park, MD, U.S.A.}
\affiliation{Institute for Research in Electronics and Applied Physics, University of Maryland, College Park, MD, U.S.A.}
\affiliation{Institute for Physical Science and Technology, University of Maryland, College Park, MD, U.S.A.}
\affiliation{Santa Fe Institute, Santa Fe, NM, U.S.A.}
\begin{abstract}
Network science is a rapidly expanding field, with a large and growing body of work on network-based dynamical processes. Most theoretical results in this area rely on the so-called \emph{locally tree-like approximation}. This is, however, usually an `uncontrolled' approximation, in the sense that the magnitudes of the error are typically unknown, although numerical results show that this error is often surprisingly small. In this paper we place this approximation on more rigorous footing by calculating the magnitude of deviations away from tree-based theories in the context of discrete-time critical network cascades with re-excitable nodes. We discuss the conditions under which tree-like approximations give good results for calculating network criticality, and also explain the reasons for deviation from this approximation, in terms of the density of certain kinds of network motifs. Using this understanding, we derive results for network criticality that apply to general networks that explicitly do not satisfy the locally tree-like approximation. In particular, we focus on the bi-parallel motif, the smallest motif relevant to the failure of a tree-based theory in this context, and we derive the corrections due to such motifs on the conditions for criticality. We verify our claims on computer-generated networks, and we confirm that our theory accurately predicts the observed deviations from criticality. Using our theory, we explain why numerical simulations often show that deviations from a tree-based theory are surprisingly small. More specifically, we show that these deviations are negligible for networks whose average degree is even modestly large compared to one, justifying why tree-based theories appear to work well for most real-world networks. 
\end{abstract}
\maketitle

\section{Introduction}\label{sec:introduction}

The study of dynamical processes on networks is among the most important areas of research in network science\cite{barrat2008dynamical,boccaletti2006complex,porter2016dynamical}. Theoretical understanding of these processes on networks found in real-world studies has wide potential impact, owing to the applicability of such systems to fields of study as diverse as epidemiology\cite{miller2009percolation, valdano2015analytical, prakash2012threshold}, neuroscience\cite{shew2009neuronal, kinouchi2006optimal, tanaka2009recurrent}, ecology\cite{cohen2012community,sole2001complexity}, electrical engineering \cite{motter2002cascade}, social sciences\cite{draief2010epidemics, iribarren2011branching} and several others. 
Obtaining rigorous analytical results for such systems on networks that are found in nature is generally very hard. At the present time a majority of central results in this area rely on the so-called `locally tree-like approximation', which neglects the effect of small loops and cycles in the network structure\cite{newman2018networks,strogatz2001exploring,dorogovtsev2008critical}. However, despite being commonly used, this approximation is uncontrolled, i.e., there do not exist clear indications of the extent of validity of the approximation, nor are estimates of the scaling of the order of error to be expected typically available. This issue is additionally compounded due to two observations: first, real-world networks tend to have a high clustering with a significant density of small loops and are hence far from tree-like\cite{newman2018networks,strogatz2001exploring}; and, second, theoretical results on the basis of locally tree-like approximations appear to be in close agreement with numerically obtained results for dynamical processes on such far from tree-like real-world networks\cite{melnik2011unreasonable}. This leads to the central foundational questions of our paper: Why do locally tree-like approximations appear to work well on real-world networks, when can we safely use these approximations, and what are the sizes of expected errors in the approximation? In this paper we work in the context of cascade processes on networks with re-excitable nodes, and aim to put the locally tree-like approximation on more rigorous footing by calculating the expected order of error in using this approximation. We also discuss the application of our results to other types of network dynamics in our discussion in Sec. \ref{sec:conclusion}

We restrict our analysis to the important dynamical process of network cascades. Cascade processes on networks, also referred to as avalanches on networks, have been widely studied due in part to their wide range of applicability, including problems relating to epidemiology\cite{miller2009percolation, valdano2015analytical, prakash2012threshold}, neuroscience\cite{shew2009neuronal, kinouchi2006optimal, larremore2011predicting, tanaka2009recurrent}, genealogy\cite{watson1875probability}, social sciences\cite{draief2010epidemics, iribarren2011branching}, and network security\cite{acemoglu2016network}. 
In a cascade process, if the average number of nodes excited by a single node is sufficiently large, then small initial perturbations can give rise to activity that may persist indefinitely. In contrast, if the average number of single-event induced excitations is too small, then cascades die out and the network can suppress the future activity resulting from even large initial perturbations. At the boundary between these two phases, the network is said to be critical. Network criticality has been studied in relation to a wide range of phenomena, such as percolation thresholds, epidemic thresholds and phase transitions in cooperative models, among others\cite{dorogovtsev2008critical}. In the particular context of neuronal networks, several studies suggest that networks of neurons tend to operate in this critical regime\cite{shew2009neuronal, poil2008avalanche, beggs2003neuronal}, which admits an increased dynamical range\cite{shew2009neuronal} and enlarged information capacity\cite{shew2011information}. Such critical phenomena are generally characterized by the presence of power-law statistics in various relevant distributions \cite{beggs2003neuronal}.

Watson and Galton\cite{watson1875probability} first studied the problem of a branching processes on a network having a tree topology in the context of the extinction of family names, and examined the case for which the `cascade' of family names would die out. Here, we consider a related problem of cascades on networks with general topologies. Motivated by the case of neuronal cascades, we specifically consider the situation in which nodes can be re-excited multiple times during the same cascade, similar to discrete SIS (susceptible-infected-susceptible) models in epidemiology (see for example Refs.\cite{shi2008sis,parshani2010epidemic,boguna2002epidemic}). 

The problem of criticality in branching processes on networks with general network topologies has been studied over the last several years, resulting in some analytical results regarding the conditions for criticality\cite{gomez2010discrete,larremore2012statistical, pastor2001epidemic, larremore2011predicting, larremore2011effects, assis2008dynamic, kinouchi2006optimal}. 
However, as discussed above, these results generally rely on `locally tree-like approximations'\cite{gomez2010discrete,larremore2012statistical,larremore2011predicting,kinouchi2006optimal}, or pertain to specific classes of networks having specific topologies\cite{moore2000epidemics,pastor2001epidemic,assis2008dynamic}. The `locally tree-like approximation'\cite{newman2018networks} is the assumption that subgraphs that extend a short distance from a given node are typically trees. For a wide range of dynamical processes on networks, the results predicted by a tree-based theory (i.e., for networks satisfying the locally tree-like approximation) have been shown to often be close to results obtained for real networks\cite{melnik2011unreasonable} that are not tree-like. 
In this paper, we provide analytic justifications for the success of the locally tree-like approximation in predicting when to expect critical cascades, analyze the reasons for the breakdown of such a tree-based theory, and develop an approach for capturing deviations from tree-like behavior. 

We use the framework previously employed by Larremore et al.\cite{larremore2012statistical} to analyze the conditions for criticality in a general network in the thermodynamic limit of large network sizes. We demonstrate that network motifs that we call $k$-parallel motifs (see Fig.\ref{fig:kparallel}) are especially relevant to the failure of tree-like approximations, and we study the corrections introduced due to such motifs. In particular, we discuss how the bi-parallel motif is the most relevant motif for deviations away from tree-like behavior, and we derive the condition for criticality for networks containing such motifs. 
Our results demonstrate why the tree-based theory always gives good results for real-world networks that are not locally tree-like and may have high clustering --- this is consistent with previous numerical studies\cite{melnik2011unreasonable}. 
Notably, we explain why critical dynamics will always be consistent with tree-like approximations in networks with even modestly large average degree. We show that when the fourth power of the average degree is large compared to 1, critical dynamics will be consistent with tree-like approximations, even when the networks is far from tree-like. Empirically, we see strong agreement with tree-like approximations in networks with average degree as small as 10.
We also consider networks constructed to have a large number of bi-parallel motifs so as to demonstrate an observable deviation from the tree-based theory for criticality and we find a close agreement of our derived results with numerical experiments. 


The remainder of this paper is organized as follows. In Sec. \ref{sec:branchingprocess} we detail the model of network cascades that we consider. In Sec. \ref{sec:treebased}, we discuss the locally tree-like approximation and briefly describe results obtained under this approximation. We then discuss the reasons for deviations away from such a theory in terms of relevant network motifs (Sec. \ref{sec:failure}), and introduce the `second-level approximation' that we develop to estimate criticality in networks that are explicitly allowed to break the locally tree-like approximation via the presence of bi-parallel motifs (Sec. \ref{sec:twoparallel}). We then present numerical results demonstrating the validity of our claims (Sec. \ref{sec:numerics}), followed by details of the derivations for the results under both the locally tree-like approximation and the second-level approximation (Sec. \ref{sec:derivations}). We conclude in Sec. \ref{sec:conclusion} with a summary of our results and a discussion of limitations of our techniques, as well as extensions to other network-based dynamical processes.

\section{Network cascade dynamics}\label{sec:branchingprocess}

Our basic setup is motivated by previous work of Larremore et al.\cite{larremore2012statistical}. We consider excitable dynamics in a network of nodes, using the following definitions:
\begin{itemize}[leftmargin=*]
	\item Network setup
	\begin{itemize}[leftmargin=*]
		\item{{$N$}             : Number of nodes in the network}
		\item{{$\widetilde{A}$} : Adjacency matrix of the network; $\widetilde{A}_{mn}=1$ if there is a directed edge from node $m$ to node $n$, and $\widetilde{A}_{mn}=0$ otherwise.}
		\item{{$\tlambda$}      : Perron-Frobenius eigenvalue of $\widetilde{A}$}
		\item{{$x_n(t)$}        : The state (either $0$ or $1$) of vertex $n$ at time $t$}
	\end{itemize}                                 
	\item Network dynamics                        
	\begin{itemize}[leftmargin=*]                 
		\item{{$p$}             : Transmission probability of excitation across any edge; constant for all edges}
		\item{{$A$}             : Probability weight matrix; $A=p\times \widetilde{A}$}
		\item{{$\lambda$}       : Perron-Frobenius eigenvalue of $A$; $\lambda=p\times\tlambda$}
		\item{{$\rho_n$}        : Duration of an avalanche starting at node $n$}
		\item{{$c_n(t)$}        : $\Pr(\rho_n \leq t)$}
		\item{{$b_n$}           : Probability that $\rho_n$ is finite; $b_n=\lim_{t\to\infty}c_n(t)$}
		\item{{$s_n$}           : Size of an avalanche starting at node $n$; $s_n=\sum_{t=0}^{\infty}\sum_{k=1}^{N} x_k(t)$}
	\end{itemize}	                                
	\item Critical network cascades               
	\begin{itemize}[leftmargin=*]                 
		\item{{$\pcrit$}        : Transmission probability such that the network exhibits criticality}
		\item{{$\lcrit$}        : Perron-Frobenius eigenvalue of $A$ such that the network exhibits criticality; $\lcrit=\pcrit\times\tlambda$}
	\end{itemize}	
\end{itemize}

We consider a directed, unweighted, strongly-connected (i.e., every node is reachable from every other node) network of $N \gg 1$ nodes, labeled by the integers from $1$ to $N$, connected according to an adjacency matrix $\widetilde{A}$. We also assume our network to have no self-loops or multiple edges. We further assume discrete time network dynamics, in which the state of the $n^{\text{th}}$ node at time $t$ is represented by $x_n(t)$, which can take a value of either $0$ or $1$. If $x_n(t)=1$ ($x_n(t)=0$) then the node $n$ is said to be in the active or excited (inactive or resting) state at time $t$. The dynamics of activation and deactivation on the network are governed by a probability weight matrix, 
\begin{equation}
A=p\times \widetilde{A},
\end{equation}
where $p$ is a transmission probability, $0<p<1$. If a node $m$ is active at time $t-1$, it sends an activating signal at time $t$ to node $n$ with probability $A_{mn}$. If, at time $t$, node $n$ receives an activating signal from any node it is activated, and all edges that have sent activating signals are said to be active at that time. An active node will relax back to the inactive state at the next time step unless it is activated again. To summarize, we can write the dynamics of cascade propagation that we consider as
\begin{equation}\label{eq:dynamics}
x_i(t) = H\left[\sum_{j=1}^N x_j(t-1) s_{ji}(t)\right],
\end{equation}
where for each $i$, $j$ and $t$, the quantity $s_{ji}(t)\in \{0,1\}$ is an independent Bernoulli random variable such that $\Pr [s_{ji}(t)=1] = A_{ji}$ and $\Pr [s_{ji}(t)=0] = 1-A_{ji}$, and $H(x)=1$ for $x>0$ and $H(x)=0$ for $x=0$. Note that in our model the state of each node is updated in parallel at each timestep.

We denote the Perron-Frobenius eigenvalue (i.e., the eigenvalue having the largest magnitude, which is unique, real, and also referred to as the leading or dominant eigenvalue) of $\widetilde{A}$ as $\tlambda$, and similarly denote the Perron-Frobenius eigenvalue of $A=p\times \widetilde{A}$ as 
\begin{equation}
\lambda = p\times \tlambda.
\end{equation}
In practice, we use $\lambda$ as the tunable parameter to vary the edge-weights $A_{mn}$, rather than directly tuning $p$.

We start from an initial condition in which all the nodes in the network are inactive, except for a single randomly chosen node, $n$ which is set to the excited state at time $t=0$: $x_n(0)=1$. The network is then allowed to evolve under the aforementioned dynamics. We refer to the sequence of resulting excited nodes as the avalanche or cascade starting at $n$.
The duration, $\rho_n$, of an avalanche starting at node $n$ is defined to be the minimum number of time steps after which all nodes are in the resting state, i.e.,
\begin{equation}
\rho_n=\min_{t\geq0} \left\{t | x_k(t)=0 \;\forall\; k \right\}.
\end{equation}
If no such minimum $t$ exists, the avalanche is said to last for an infinite duration. The size of an avalanche starting at node $n$ is defined as 
\begin{equation}
s_n = \sum_{t=0}^{\infty}\sum_{k=1}^{N} x_k(t).
\end{equation}
Note this is not the number of nodes involved in the avalanche, but rather the number of node activations, and a single node may be activated multiple times during one avalanche. The size of avalanches lasting for infinite duration is therefore infinite. Due to the probabilistic nature of activity propagation across edges, $\rho_n$ and $s_n$ are both random variables whose distributions characterize criticality in networks.

A cumulative distribution function of avalanche durations can be defined as 
\begin{equation}
c_n(t)=\Pr(\rho_n \leq t),
\end{equation}
which is the probability that an avalanche that begins at node $n$ has a duration that is less than or equal to $t$. Note that from the initial conditions in the problem, $c_n(0)=0$ for all $n$. Further, since $c_n(t)$ is necessarily a non-decreasing function that is bounded above by $1$, it must converge to some limit $b_n = \lim_{t\to \infty} c_n(t)$, which is the probability that an avalanche starting from node $n$ has a finite duration.  
In particular, for small values of $p$, all avalanches eventually subside, and hence have a finite duration, giving $b_n=1$. Such networks are called \emph{subcritical}. For sufficiently large values of $p$, there is a positive fraction of avalanches that last for an infinite duration and hence $b_n<1$. Such networks are called \emph{supercritical} networks. As we increase $p$ to go from a subcritical network to a supercritical network, there is a transition between the two phases, corresponding to the largest value of $p$ such that $b_n=1$. 
At this `critical' transition probability, denoted by $\pcrit$, the network is said to be in the \emph{critical} state, or equivalently, the network is said to exhibit \emph{criticality}. 

This phase transition is demonstrated in Fig. \ref{fig:phasetransition} for a variety of different networks, each with approximately $N=4\times10^4$ nodes (for the algorithms used to construct these networks, and to demonstrate that these networks are sufficiently large to study criticality, see Appendix \ref{apx:bigenough}). 
For each network we consider $10^6$ avalanches initialized at random nodes, and measure the fraction of avalanches that last for durations longer than $N/15$ timesteps. Since we consider the avalanches initialized in this random fashion, this fraction can be interpreted as the average value of $(1-b_n)$ over all nodes, which we represent as $(1-b)$. Note how $(1-b)$ remains close to zero  for low values of Perron-Frobenius eigenvalue, $\lambda$, until a critical phase transition, after which it rapidly increases.
We denote the Perron-Frobenius eigenvalue of the critical network $A=\pcrit\times \tA$ as $\lcrit$.
We are interested in estimating this quantity, and in Sec. \ref{sec:numerics} we compare the estimates according a tree-based theory with our `second-level approximation'  to empirically determined values for $\pcrit$ and $\lcrit$. Note that our work does not address the distributions of cascades away from criticality. While Ref. \cite{melnik2011unreasonable} discusses numerical results for network cascades both at criticality and away from criticality, our work will only address theoretical results for estimating $\lambda_c$, corresponding to networks at criticality.

We use the $\kappa$-metric developed in Ref.\cite{shew2009neuronal} to characterize whether or not a given network is displaying criticality (For completeness, Appendix \ref{apx:kappadefinition} gives the definition of $\kappa$). Since criticality is often characterized by power-law distributions of the avalanche sizes and the avalanche durations\cite{shew2009neuronal,beggs2003neuronal}, this $\kappa$-metric was introduced to characterize the deviations of a given distribution of avalanche sizes away from the power-law fit to the data. 
$\kappa=1$ indicates a network at criticality, corresponding to an observed distribution of avalanche sizes close to a power-law distribution. $\kappa<1$ indicates subcritical network avalanches, with the distribution of avalanche sizes exponentially decaying at large duration values, and $\kappa>1$ indicates supercritical network avalanches, with a measurable fraction of avalanches  that are infinitely long. Examples of such distributions are shown in Fig. \ref{fig:threecurves}, where we present log-log plots of the Complementary Cumulative Distribution Functions for cascade sizes in each of the three regimes (i.e., the probability that a cascade starting with a random initial perturbation has a size less than or equal to $s$, as a function of $s$) in the subcritical, critical and supercritical regimes.

\subsection{Locally tree-like approximation}\label{sec:treebased}
We say that a network satisfies the locally tree-like approximation if, for any two nodes $m$ and $n$ that are separated by a path with a small number of edges $l_1 \ll N$, the probability that there exists another path from $m$ to $n$ of length $l_2 \leq l_1$ is negligible.

Under this approximation we assume that events occurring along different edges that lead away from the same node are independent from each other, since the approximation implies that the corresponding activation paths do not share common nodes. Using this approximation, the following recursion relation for $c_n(t)$ holds:
\begin{equation}\label{eq:original}
c_n(t)=\prod_m [(1-A_{nm})+A_{nm} c_m(t-1)].
\end{equation}
This recursion relation was used by Larremore et al.\cite{larremore2012statistical} to demonstrate that the network exhibits criticality when the Perron-Frobenius eigenvalue of the matrix $A$ is one, i.e.,
\begin{equation}\label{eq:pcritonebylambda}
\lcrit = 1 \text{   or   } \pcrit = (\tlambda)^{-1},
\end{equation}
where $\tlambda$ is the Perron-Frobenius eigenvalue of the adjacency matrix $\widetilde{A}$. This can also be seen in Fig. \ref{fig:phasetransition}, where the phase transition to criticality for locally tree-like networks, such as a large random Erd\H{o}s-R\'{e}nyi network (shown in red squares) appears to occur at $\lambda=1$.


\subsection{Reasons for success and conditions for failure of the tree-based theory}\label{sec:failure}
\begin{figure}
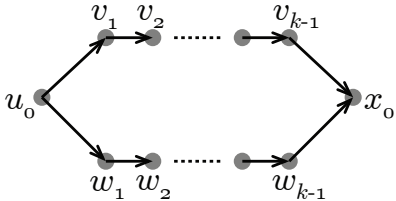

\includegraphics[width=0.6\columnwidth]{{{kparallel}}}
\caption{The structure of the $k$-parallel motif, which is relevant to the deviations of the branching process dynamics from a tree-based theory. Here, $u_0$ is the initial node, $x_0$ is the final node, and the paths $u_0\to v_0 \to \cdots v_{k-1} \to x_0$ and $u_0\to w_0 \to \cdots w_{k-1} \to x_0$ form the two distinct $k$ length paths that are used to generate this motif.
}
\label{fig:kparallel}
\end{figure}

In order to derive the results in Eq. (\ref{eq:pcritonebylambda}), the main assumption made about the network structure is the locally tree-like approximation, which is manifested in Eq. (\ref{eq:original}). In order to calculate $c_n(t)$, Eq. (\ref{eq:original}), the probability distributions $c_m(t-1)$ are assumed to be independent distributions for each node $m$. This assumption is equivalent to assuming that the cascades of excitation propagating from different edges starting from the same node do not interact. Let us assume that this were not the case, and that cascades resulting from different edge excitations do interact. Note that under the dynamics that have been defined, two excitation cascades starting at a node $n$ can only interact at time $k$ if some node $m$ is being excited by activation of two different edges simultaneously. This would mean that there exist two $k$-length paths between the nodes $n$ and $m$. We shall call the motif generated by the two distinct paths of length $k$ from the same initial node to the same final node as shown in Fig. \ref{fig:kparallel} to be $k$\emph{-parallel}. (This nomenclature has been chosen to be similar to earlier nomenclature of the \emph{bi-parallel} motif, such as in Ref.\cite{milo2002network}, which is then equivalent to what we term the $2$-parallel motif.) Thus we see that in the described dynamics, the motifs that are primarily responsible for deviations from the tree-based theory are $k$-parallel motifs, the smallest such motif being the $2$-parallel motif. 

For the case of epidemiological models involving SIR-like dynamics (for a description of SIR dynamics see Refs.\cite{volz2008sir,miller2011note,may2001infection}), i.e., dynamics in which each node is activated exactly once and is then removed from the network, the smallest motif that allows interaction between cascades is the feed-forward triangular motif. Since our consideration is restricted to SIS-like dynamics, i.e., dynamics in which each node may be activated repeatedly, such triangular motifs are not relevant to the breakdown of the tree-based theory. 

Thus, we note that while SIR and SIS dynamics are identical on directed networks near the critical point under the locally tree-like approximation, they can be distinguished from each other once this approximation is invalidated. (For recent work analyzing SIR-like dynamics on networks beyond the locally tree-like approximation see Ref.\cite{radicchi2016beyond}.) 
For additional discussion regarding SIR-like dynamics on networks see Sec. \ref{sec:conclusion}

Note that while these statements are strictly true in our model described in Sec. \ref{sec:branchingprocess}, there are some caveats that must be taken into account to make a similar statement applying to real-world scenarios. The model that we have described is a discrete state and discrete time model. For our model to be a reasonable representation of a real-world continuous time model, we are effectively making the assumption that the time-scale of propagation of node activity through each edge is approximately the same for all edges, and similarly the time-scale for which a node remains active before relaxing back into an inactive state is also approximately the same for all nodes. Further, in our discrete time model we have assumed that nodal state updates occur approximately synchronously for all nodes in the network. 

\begin{figure}
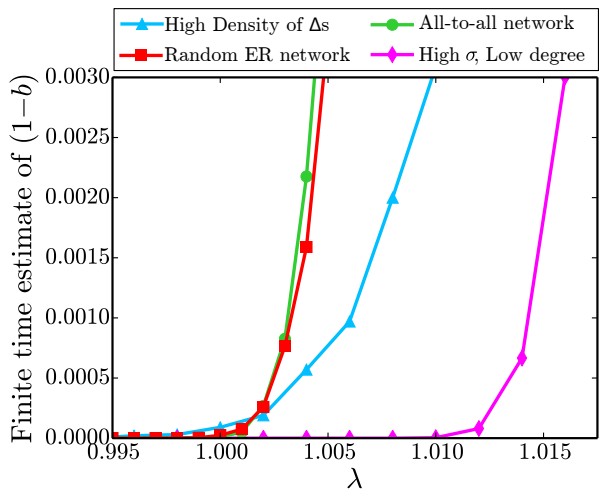

\includegraphics[width=\columnwidth]{{{Phase_transitions_w_poissonian_new_updated_legend}}}
\caption{Phase transition in the network activity $(1-b)$ vs the leading eigenvalue $\lambda$ of the weighted adjacency matrix for different types of networks. Each network was generated with $N_0=5\times 10^5$, corresponding to the algorithms given in Appendix \ref{apx:bigenough}. Note that networks that have a high clustering coefficient, i.e., have a high density of triangle motifs, (shown in blue triangles; data generated for a network with $N\approx 4.7\times 10^4$ nodes with average degree of $3.5$), random Erd\H{o}s-R\'{e}nyi networks (shown in red squares; data generated for a network with $N\approx 3.1\times 10^4$ nodes with average degree of $3.3$) and all-to-all networks (shown in green circles; data generated for a network with $N=5\times 10^4$ nodes) all exhibit the critical phase transition at $\lambda=1$, as predicted by a tree-based theory (see Eq. \ref{eq:pcritonebylambda}). Networks with a high density of 2-parallel motifs ($\sigma\approx0.04$, where $\sigma$ defined in Eq. (\ref{eq:sigma}) is a normalized parameter measuring the density of $2$-parallel motifs in the network) and a low average degree ($\approx 3.3$ in this case) exhibit a phase transition at a distinctly larger value of $\lambda$ (shown in the magenta diamonds; network constructed following the algorithm in Appendix \ref{apx:algoER}, with a network size of $N=3.5\times 10^4$ nodes). Note that all-to-all networks also have a high density of 2-parallel motifs, \emph{but} they have a very large average degree, and hence continue to exhibit a phase transition at $\lambda=1$ (see Sec. \ref{sec:failure} for more details).}
\label{fig:phasetransition}
\end{figure}

As discussed above, triangular motifs are not directly relevant to the deviations from the tree-based theory for the dynamics we study, which is consistent with the observations Melnik et al.\cite{melnik2011unreasonable}, where the authors note that clustering coefficients are not highly relevant to the deviations from the tree-based theory. To test the hypothesis that $k$-parallel motifs are essential as opposed to triangular motifs, for deviations from the locally tree-like approximation, we construct networks by randomly choosing triples of nodes that are connected as triangular feed-forward motifs. 
These networks have a low number of $k$-parallel motifs. We then set $p$ such that $\lambda=1$ and numerically simulate $10^7$ avalanches with random initially activated nodes. We obtain a value of $\kappa=1-1.3\times 10^{-3} \approx 1$, indicating that the prediction according to the locally tree-like approximation, i.e., $\lambda=\lcrit=1$, accurately predicts criticality in the network. This can also be seen in Fig. \ref{fig:phasetransition}, where the phase transition to criticality for such a network (shown in blue triangular markers) appears to occur at $\lambda=1$. 
We observe a similar behavior for triplets of nodes connected in cyclic motifs instead of feed-forward motifs.

Importantly, we also note that the presence of a large number of $k$-parallel motifs does not guarantee large deviations from the tree-based theory. To demonstrate this, we use the example of a network with all-to-all connections. By construction, all-to-all networks have the maximum possible number number of $k$-parallel motifs for each $k$ at each node. Numerical simulations of cascades on all-to-all networks with $N=4\times 10^4$ nodes with $p$ such that $\lambda=1$ yielded $\kappa=1-1.4\times10^{-3}\approx 1$, indicating criticality and hence demonstrating no significant deviation from the prediction according to the locally tree-like approximation. This can also be seen in Fig. \ref{fig:phasetransition}, where the phase transition to criticality for an all-to-all network (shown in green circular markers) appears to occur at $\lambda=1$.
To see why this is the case, observe that in an all-to-all network, the number of paths of length $k$ between two given nodes scales as $N^{k-1}$, whereas the total number of paths of length $k$ starting at a given node scales as $N^k$. Thus the probability for two excitation cascades to meet at any given node scales as $N\times (N^{k-1}/N^k)^2 \sim 1/N$ and hence tends to zero as $N$ goes to $\infty$.

\begin{figure*}
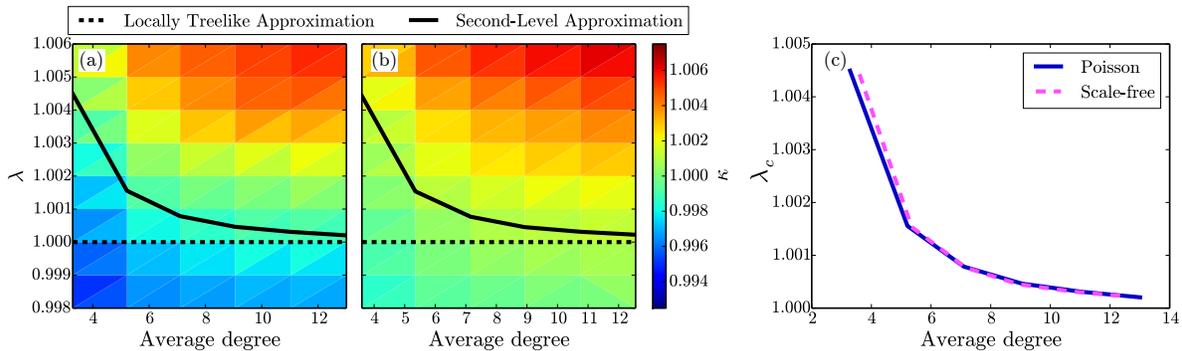

\includegraphics[width=1.8\columnwidth]{{{varying_degree_both_w_comp}}}
\caption{The $\kappa$ measure for criticality for networks in the phase space of average degree versus $\lambda$, for an approximately constant density of $2$-parallel motifs for networks with: (a) a Poisson degree distribution, and (b) a scale-free degree distribution. These networks were generated using the algorithms described in Appendix \ref{apx:algorithms}. To ensure a constant $\sigma\approx0.04$ across all networks, pairs of edges in networks with larger $\sigma$ were randomly swapped to maintain the joint in-out degree distribution while reducing $\sigma$ to the lowest $\sigma$ among all generated networks. The dashed black curve corresponds to the prediction for network criticality according to a tree-based theory, and the solid black curve corresponds to the prediction according to the second-level approximation derived in this paper. 
Note that the empirical observation of criticality (corresponding to the green (light gray) region with $\kappa\approx1$) as well as the prediction using our second-level approximation equations rapidly approach the prediction according to a tree-based theory with increasing average network degree. We expect improving agreement between the empirical predictions and our second-level approximation with increasing network sizes. For low average degree and low $\lambda$ (lower left), $\kappa$ is below 1 (blue); for high average degree and high $\lambda$ (upper right), $\kappa$ is above 1 (red). Also note that the green (light gray) region for the case of the scale-free degree distribution is significantly broader, possibly related to previous observations in Refs. \cite{callaway2000network,albert2000error} regarding the robustness of scale-free networks in percolation-like problems. The prediction according to the derived second-level approximations are overlayed on each other in (c) to demonstrate that the magnitude of $\lcrit$ does not strongly depend on the network topology, and decays rapidly towards one with increasing average degree in both cases.
}
\label{fig:varyingd}
\end{figure*}

In general, the deviations from a tree-based theory are suppressed by large average degree in the network for similar reasons, since it is increasingly unlikely for cascades beginning at a given node to interact at a later time. This is demonstrated in Fig. \ref{fig:varyingd}, which shows a comparison between our theoretical results for $\lcrit$ (shown in the black solid curve) with numerical estimates for $\lcrit$ using the $\kappa$ metric ($\kappa$ shown in color; the numerical estimate for $\lcrit$ corresponds to the green (light gray) region representing $\kappa\approx 1$) for networks with varying average degree with a constant density of 2-parallel motifs. Note that both the numerical estimate as well as our theoretical result rapidly approach 1 with increasing average degree. Particularly, average degrees as small as 10 are sufficient for the deviations away from a tree-based theory to be negligible. We further quantify this intuition in Sec. \ref{sec:pcrit}. In Fig. \ref{fig:phasetransition} we see that all-to-all networks (which have the maximum possible density of 2-parallel motifs with the maximum possible average degree) continue to show a phase transition at approximately $\lambda=1$, whereas a high density of 2-parallel motifs in a network with a low average degree of 3.3 (magenta diamond markers) shows a phase transition at a significantly larger value of $\lambda$ (to demonstrate that the networks we studied are sufficiently large to study criticality, see Appendix \ref{apx:bigenough}).

In addition, we note that the effect of interaction between two excitation paths is always to reduce the number of active nodes at any given time with respect to the number of active nodes expected according to a tree-based theory. This is because interaction effectively makes the transmission of an excitation through an edge immaterial if the corresponding node is excited by another node. The probability of an avalanche to last for any given duration is less due to the presence of such an interaction, and the addition of interaction effectively suppresses the overall number of transmitted excitations. Hence, a tree-based theory always under-predicts the value $\pcrit$ with respect to the actual critical transmission probability, i.e., for networks that may not satisfy the locally tree-like approximation, $(\tlambda)^{-1} \leq \pcrit$ or $\lcrit \geq 1$. We verify this numerically in Sec. \ref{sec:numerics}.

Previous work\cite{melnik2011unreasonable} demonstrated that shorter mean inter-vertex length correlates well with smaller deviations from the results of a tree-based theory (in particular, the difference between the mean inter-vertex length for the given network and the mean inter-vertex length for a random network with the same size and joint degree distribution as the given network). Our interpretation of this result is as follows: For networks with a low mean inter-vertex length, $l$, the number of nodes that can be reached from any given node in a short number of steps rises very rapidly with number of steps. Assume that in $\scO(l)$ number of steps, the number of nodes reached is $\scO(N)$. Thus if we consider two random paths of the same length $k$ from the same starting node, the probability that they both end at the same node is very small. In particular, if $k\sim \scO(l)$, then the probability that a $k$-parallel motif might be relevant to the interaction of two excitation paths scales as $\scO(1/N)$. 
Thus, a network with a small mean inter-vertex length would be expected to have a low density of $k$-parallel motifs for small $k$. Following our earlier discussion on the importance of $k$-parallel motifs to deviations away from the locally tree-like approximation, we note that our result is hence consistent with the observations in Ref. \cite{melnik2011unreasonable}.

In any iteration, the probability of a given $k$-parallel motif being relevant to such interactions scales as $p^{2k}$, and hence only $k$-parallel motifs for small $k$ are relevant to the deviations from a tree-based theory (since $0<p<1$). Under the assumption of no double edges in the network, the smallest $k$-parallel motif is the $2$-parallel motif. Hence in what follows we neglect the effect of $k$-parallel motifs with $k>2$ and focus on the correction to an estimate for $\pcrit$ due to the presence of $2$-parallel motifs. We note that in a similar fashion to the discussion above, ignoring the interaction between excitation cascades due to $k$-parallel motifs for $k>2$ leads to an underestimate of the transmission probability $p$ for criticality.


\section{Corrections due to 2-parallel motifs}\label{sec:twoparallel}

According to the tree-based theory, in Eq. (\ref{eq:original}), when calculating the cumulative distribution at a node, we assume that the cumulative distributions at time $(t-1)$ are all independent of one another. We now consider what happens if we break this assumption, and assume instead that while distributions at time $(t-1)$ may not be independent of one another, distributions at time $(t-2)$ are independent of one another. We refer to results obtained assuming this condition as the `second-level approximation'. This approximation only takes into account the effect of $2$-parallel motifs. While we present the details and derivations of our second-level approximation in Sec. \ref{sec:derivations}, a short summary of our argument is as follows:
\begin{enumerate}
	\item Note that the product in Eq. (\ref{eq:original}) explicitly assumes the locally tree-like approximation. This product can be rewritten as a summation (Eq.(\ref{eq:firstlevelrecursion})) where an important quantity in the expression is the number of nodes reached from paths of lengths 1 and 2 starting at any node. In this summation form of the expression, the locally tree-like approximation is now encoded in writing this number of nodes reached in terms of the degree distributions of the network.
	\item For networks that explicitly do not satisfy the locally tree-like approximation, we write a variant of the summation form Eq. (\ref{eq:original}) where the relevant quantity in the expression is again the number of nodes reached (Eq. (\ref{eq:secondlevel})) --- however, when the locally tree-like approximation does not hold, the number of nodes reached does not depend only on the degree distribution, but also on the density of small motifs. As discussed earlier, the smallest relevant motif is the $2$-parallel motif, hence we calculate the average number of nodes reached by paths of length 1 and 2 in terms of the degree distribution as well as the density of $2$-parallel motifs.
	\item As discussed in Sec. \ref{sec:branchingprocess}, for critical cascades the probability that cascades will not persist indefinitely, $b$, is equal to one. Following the analysis in Ref. \cite{larremore2012statistical}, it can be shown that at criticality $b=1$ is in fact a double root of Eq. (\ref{eq:original}) in the limit of $t\to\infty$ (In this limit note that $b=1$ is always a solution, however at criticality an additional root at $b=1$ appears, resulting in the double root). We use a similar double root criteria on Eq. (\ref{eq:secondlevel}) to obtain an expression relating the probabilities of activity transmission across edges in the network to quantities depending on the degree distribution \emph{and} the density of $2$-parallel motifs (Eq. (\ref{eq:secondlevelpcrit}), and in Appendix \ref{apx:cutoffexpressions}).
\end{enumerate}

\section{Numerical verification of results}\label{sec:numerics}

Our second level prediction for $\pcrit$ is derived in Sec. \ref{sec:derivations} and given in Eq. (\ref{eq:secondlevelpcrit}).
To test this prediction we generate networks designed to exhibit a large deviation from a tree-based theory by growing networks with a high density of $2$-parallel motifs. In Appendix \ref{apx:algorithms} we describe the algorithms used to construct such networks. In particular, Appendix \ref{apx:algoscalefree} describes the construction of a network with a scale-free degree distribution, and Appendix \ref{apx:algoER} describes the construction of a network with a sharper, approximately Poisson degree distribution. Fig. \ref{fig:smallnetwork} shows an example of a small network generated via the algorithm described in Appendix \ref{apx:algoER}.

\begin{figure}
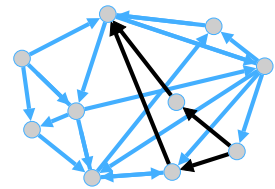

\includegraphics[width=0.4\columnwidth]{{{small_sharp_graph}}}
\caption{Example of a small network generated via the algorithm described in Appendix \ref{apx:algoER} with $N_0=15$. An example of a 2-parallel motif in the network is shown in black.
}
\label{fig:smallnetwork}
\end{figure}

Unless otherwise specified, in all cases we consider network with approximately $5\times10^4$ nodes with algorithm parameters chosen as described in the Appendices to result in an average degree of about $3.5$. We numerically simulate $10^6$ iterations of the branching process dynamics as described in Sec. \ref{sec:branchingprocess} to obtain avalanche distributions for varying values of $\lambda$. As discussed earlier in Sec. \ref{sec:branchingprocess}, critical avalanches are characterized by power-law statistics, and we use the $\kappa$ metric to evaluate whether the obtained avalanche distributions correspond to criticality. We use the Python package `\texttt{powerlaw}' created by Alstott et al.\cite{alstott2014powerlaw}, which uses the tools developed by Clauset et al.\cite{clauset2009power} and Klaus et al.\cite{klaus2011statistical} to calculate the value of $\kappa$ for the obtained distributions for each value of $\lambda$. 

\begin{figure}
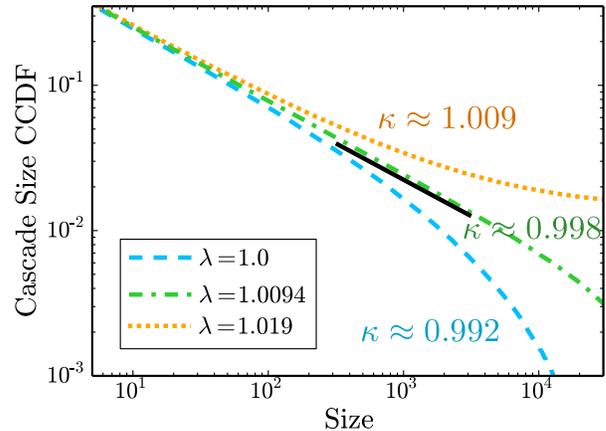

\includegraphics[width=\columnwidth]{{{3_distributions_sharp_dist_theory_lambda}}}
\caption{Complementary Cumulative Distribution Functions (CCDF) of avalanche sizes on a network with a large number of $2$-parallel motifs (according to the algorithm described in Appendix \ref{apx:algoER} with $N_0=5\times 10^4$) for $\lambda=1.0$, $1.0094$, and $1.019$. Note that the curve corresponding to $\lambda=1$ (shown in the blue dashed curve) does not correspond to criticality (as indicated by $\kappa<1$), and corresponds to a subcritical network. At $\lambda=1.0094$ (shown in the green dash-dotted curve) the network appears to be closest to criticality (as indicated by $\kappa \approx 1$) and for a larger value of $\lambda$, such as $1.019$ (shown in the orange dotted curve), the network appears to be supercritical (as indicated by $\kappa>1$). The black solid line is a line of slope $-0.5$, indicative of the expected exponent of the power-law distribution of avalanche sizes according to a tree-based theory (the vertical position of the black solid line is arbitrary). 
}
\label{fig:threecurves}
\end{figure}

To demonstrate the shift in criticality due to the presence of 2-parallel motifs, we present representative distributions of avalanche sizes on one such generated network. We calculate the complementary cumulative distribution function (CCDF) for the distribution of avalanche sizes, i.e., the probability that an avalanche has a size greater than or equal to a given size $s$, as a function of $s$, $[1-\langle \Pr(s_n \leq s)\rangle_n]$, where $\langle\hdots\rangle_n$ denotes an average over $n$. We present log-log plots of this CCDF in Fig. \ref{fig:threecurves}, for three values of $p$, corresponding to Perron-Frobenius eigenvalues of $\lambda=1.0$ (the critical value of $\lambda$ predicted by the locally tree-like approximation), $\lambda=1.0094$ (the critical value predicted by our second level approximation Eq. (\ref{eq:secondlevelpcrit})) and $1.019$. 
Consistent with our theoretical analysis, networks with $\lambda=1$ are seen to be subcritical, while our second level approximation, $\lambda=1.0094$, yields the closest correspondence to criticality, and $\lambda=1.019$ is supercritical. 
The $\kappa$ values corresponding to these three curves are $0.992$, $0.998$, and $1.009$, respectively.

We note that the $\lambda=1.0094$ curve is almost perfectly linear out to a size of $10^4$ and only noticeably  begins to deviate away from linearity as the size nears $10^5$ where the curve starts to have discernible downward curvature. This indicates a slight degree of subcriticality (reflected by the value of $\kappa$, $\kappa = 0.998 < 1$). This behavior is to be expected, since, as discussed subsequently, our ``second level'' theory only takes into account 2-parallel motifs, while neglecting the effect of $k$-parallel motifs (Fig. \ref{fig:kparallel}) for $k>1$, and thus is expected to slightly underestimate the critical value of $\lambda$. For reference, numerically stepping through values of $\lambda$ indicates that $\kappa=1$ at an eigenvalue of $\lambda\approx 1.011$.

\begin{figure*}
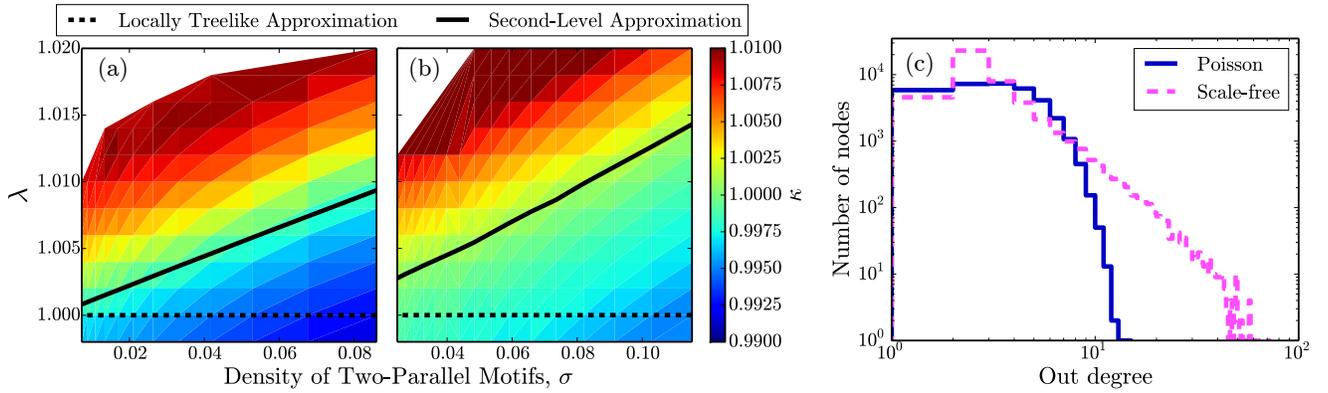

\includegraphics[width=2\columnwidth]{{{lambda_vs_sigma_N50000_both_w_dists}}}
\caption{Comparison of the estimated value of $\lcrit$ as determined empirically [green (light gray) region of plots, corresponding to $\kappa\approx 1$], with the prediction according to a tree-based theory (black-dashed line) and the prediction according to the second-level approximation equations, using a cut-off of $\alpha_{thr}=6$ (black solid line) for a network with varying densities of 2-parallel motifs as measured by the quantity $\sigma$. The network corresponding to the largest shown values of $\sigma$ is generated to have a Poisson degree distribution in (a) (following the algorithm described in Appendix \ref{apx:algoER}), and a scale-free degree distribution in (b) (following the algorithm described in Appendix \ref{apx:algoscalefree}). The networks used in (a) have $34710$ nodes and an average degree of 3.3; and networks used in (b) have $47725$ nodes with an average degree of 3.6 (Since the last step of the algorithms involve taking the strongly connected component of the network we cannot precisely tune the number of nodes and average degree of these networks easily). The out-degree distributions of the two networks are shown in (c). As described in the main text, the variation in $\sigma$ across the networks is induced by swapping increasing numbers of randomly chosen edges in a network to maintain the same degree distribution while decreasing the number of 2-parallel motifs. Note the significantly better prediction for criticality following our second-level approximation as compared with the locally tree-like approximation. (The white regions in (a) and (b) correspond to cases wherein numerical data was not generated due to large run-times associated with the highly super-critical networks) For a high density of 2-parallel motifs and low $\lambda$ (lower right), $\kappa$ is below 1 (blue); for a low density of 2-parallel motifs and high $\lambda$ (upper left), $\kappa$ is above 1 (red).
}
\label{fig:swaps}
\end{figure*}

To test the effect of 2-parallel motifs on deviations from a tree-based theory we start with a network with a relatively high density of 2-parallel motifs, and then generate a set of networks having the same joint in-out degree distribution by swapping the destination nodes of a collection of randomly chosen edge pairs. By increasing the number of swapped edge pairs, we generate networks that have a decreasing density of $2$-parallel motifs, ranging from a maximum in the initial network, to close to zero for a network in which a very large number of edges have been swapped. 
For each such network, we vary the excitation transmission probability across edges by varying the Perron-Frobenius eigenvalue $\lambda$ of the weighted adjacency matrices of the networks. At each eigenvalue, we numerically simulate $10^6$ cascades following the dynamics described in Sec. \ref{sec:branchingprocess} and calculate $\kappa$ based on the distribution of avalanche sizes. Values of $\kappa$ closer to $1$ indicating criticality are represented by the green regions in Fig. \ref{fig:swaps}. We compare this to the value of $\lcrit$ as predicted by a tree-based theory (described in Sec. \ref{sec:treebased}, results proven in Sec. \ref{sec:pcrittreelike}) and the prediction made by the analysis using the second-level approximation equations (described in Sec. \ref{sec:twoparallel} and derived in Sec. \ref{sec:pcritsecondlevel}). We present the results of this comparison in Fig. \ref{fig:swaps}, where we plot the results as a function of the density of 2-parallel motifs, $\sigma$ (defined in Eq. (\ref{eq:sigma})). For a locally tree-like network, without the presence of any 2-parallel motifs, $\sigma=0$, and for an all-to-all network, with each node containing the maximum possible number of two parallel motifs, $\sigma=1$.  We see that the prediction for $\lcrit$ according to the locally tree-like approximation (dashed black line) is always less than or equal to the observed values of $\lcrit$, shown in green. Further, we see that in all cases, the second-level approximation (solid black line) is significantly better than the tree-based theory at predicting the observed values of $\lcrit$.

\section{Derivations}\label{sec:derivations}

Our proof is structured according to the summary presented in Sec. \ref{sec:twoparallel}. In this section and in the Appendices, apart from the definitions made earlier in Sec. \ref{sec:branchingprocess}, we also make use of the definitions presented below:
\begin{itemize}[leftmargin=*]
    \item{{$\mathcal{V}$					} 	: Set of vertices in the network}
    \item{{$\mathcal{E}$					} 	: The set of the pairs of vertices corresponding to directed edges in the network; $(m,n)\in \mathcal{E} \iff \tA_{mn}=1$}
    \item{{$d(n)$									}	: Out degree of node $n$}
    \item{{$v^{PF}$								}	: Perron-Frobenius eigenvector of $\widetilde{A}^T$}
    \item{{$\dinf$						    } 	: Average degree of nodes weighted by $v^{PF}$}
    \item{{$n_0$									} 	: Initial node that we consider}
    \item{{$n_j$									} 	: Variable that spans over nodes at a distance $j$ from node $n_0$}
    \item{{$\mathcal{E}^0_k$			} 	: A $k$ element subset of $\mathcal{E}$ such that the edges in the set start at node $n_0$}
    \item{{$\mathcal{V}^0_k$			} 	: Set of vertices that the edges in $\mathcal{E}^0_k$ point to; $|\mathcal{V}^0_k| = |\mathcal{E}^0_k|$}
    \item{{$\mathcal{E}^1_r$			} 	: An $r$ element subset of $\mathcal{E}$ such that the edges in the set start at a node in $\mathcal{V}^0_k$}
    \item{{$\mathcal{V}^1_r$			} 	: Set of vertices that the edges in $\mathcal{E}^1_r$ point to; $|\mathcal{V}^1_r| \leq |\mathcal{E}^1_r|$}
    \item{{$\delta r$							}	: $|\mathcal{E}^1_r| - |\mathcal{V}^1_r| = r - |\mathcal{V}^1_r|$}
    \item{{$D[\mathcal{V}^0_k]$		}	: Sum of degrees of nodes in set $\mathcal{V}^0_k$}
    \item{{$Q_{\alpha}(n)$				} 	: The number of nodes that are reachable from node $n$ via exactly $\alpha$ edge-independent paths of length $2$}
    \item{{$\overline{Q_{\alpha}}$}	: Average of $Q_{\alpha}(n)$ over all nodes weighted by $v^{PF}$}
    \item{{$\sigma$								}	: A single parameter of the network defined to count the normalized effect of all $\overline{Q_{\alpha}}$s; defined in Eq. (\ref{eq:sigma})}
		
\end{itemize}

\subsection{Estimating the critical transition probability using the locally tree-like approximation}\label{sec:pcrittreelike}

To derive an explicit expression for the critical transition probability, $\pcrit$, we restrict ourselves to networks that are near criticality, i.e., networks for which $p$ is approximately the same as $\pcrit$. This condition implies that $b_n$ (the probability that an avalanche starting at node $n$ is finite) and correspondingly, $c_n(t)$ (the probability that an avalanche has duration less than $t$) for large $t$, are both close to $1$, and hence also close to each other. 

Thus we ignore differences between $c_n(t)$ for various $n$ when compared with quantities comparable to $1$. 
We rewrite Eq. (\ref{eq:original}) as 
\begin{equation}
c_{n_0}(t) = \prod_{(n_0,n_1) \in \mathcal{E}} [(1-p)+p c_{n_1}(t-1)],
\end{equation}
where $\mathcal{E}$ is the edge set of the adjacency matrix $\widetilde{A}$ (i.e., $(m,n)\in \mathcal{E}$ if and only if $\widetilde{A}_{mn}=1$) and hence $n_1$ spans over all nodes that have a directed edge from $n_0$ to $n_1$. More generally, we will use the index $n_k$ to span over all nodes that are at a distance of exactly $k$ steps away from $n_0$. Writing the out-degree of a node $n_0$ as $d(n_0)$, we can approximate the above expression and rewrite it as  
\begin{equation}\label{eq:firstlevelrecursion}
0= c_{n_0}(t) - \left[ (1-p)+p \frac{1}{d(n_0)}\sum_{(n_0,n_1) \in \mathcal{E}} c_{n_1}(t-1)\right]^{d(n_0)}.
\end{equation}
The approximation is valid since for any set of arbitrary quantities $X_i$ that are close to each other
\begin{align}
\prod^m_{i=1} X_i & =  \langle X\rangle^m \left(1+ \frac{1}{\langle X\rangle}\sum \delta X_i + ... \right), \\
                  & \approx \langle X \rangle^m,
\end{align}
when $m \times \delta X_i =m\times ( X_i - \langle X \rangle)\ll \langle X \rangle$.

We can then Eq. (\ref{eq:firstlevelrecursion}) as a recursion relation to write $0=c_{n_0}(t) - f(c_{n_k}(t-k)) = 0$ in terms of a known function $f$, for some $t\gg k \gg 1$. We then take the limit $t\to\infty$, and the limit $k\to\infty$. Note that since Eq. (\ref{eq:firstlevelrecursion}) is true for all times and at all nodes, the two limits can be interchanged. In taking the limit of $t\to \infty$, we have an equation of the form $0=F(\{b_i\})$. As discussed in Sec. \ref{sec:branchingprocess}, for networks that are critical or subcritical, $b_n=1$ for each $n$, and for networks that are supercritical, $b_n<1$ for each $n$. It should be noted that $b_n=1$ for each $n$ is always be a solution for the equation $0=F(\{b_i\})$, since setting $c_n(t)=1$ for each $n$ is always a solution to Eq. (\ref{eq:original}). In general, from arguments given in Ref. \cite{larremore2012statistical}, we note that there are always two solutions for $b_n$. One solution of $0=F({b_i})$ is at $b_n=1$. For supercritical networks, the other solution is less than 1, and is the value of $b_n$ to which $c_n(t)$ converges.
For subcritical networks, the other solution is greater than 1, and hence $c_n(t)$ converges to the solution at $1$ in this case instead. Since we are interested in networks at criticality, we are interested in calculating the conditions for a degeneracy of the two roots of $b_n$ at $b_n=1$ at each $n$.  We do this by looking at the subspace of $b_n=b$ for all $n$, and by solving for the vanishing derivative of $F$ with respect to $b$ at $b=1$, which results from the existence of the double root. We first take the derivative on Eq. (\ref{eq:firstlevelrecursion}), and then, after taking the appropriate limits, use the result to derive the condition on $p$ for the vanishing derivative for $F$ at $b=1$:
\begin{widetext}
\begin{align}
0 & = \frac{\partial c_{n_0}(t)}{\partial b} - \sum_{(n_0,n_1) \in \mathcal{E}} p \times \frac{\partial c_{n_1}(t-1)}{\partial b} \times \left[ (1-p)+p \frac{1}{d(n_0)}\sum_{(n_0,n_1) \in \mathcal{E}} c_{n_1}(t-1)\right]^{d(n_0) -1}, \nonumber \\
  & = \frac{\partial c_{n_0}(t)}{\partial b} - \sum_{(n_0,n_1) \in \mathcal{E}} \; \sum_{(n_1,n_2) \in \mathcal{E}} p^2 \times \frac{\partial c_{n_2}(t-2)}{\partial b} \times \xi(t|1) \times \xi(t|2), \nonumber \\
  & \vdots  \nonumber \\
  & = \frac{\partial c_{n_0}(t)}{\partial b} - \sum_{(n_0,n_1) \in \mathcal{E}} \hdots \sum_{(n_{(k-1)},n_k) \in \mathcal{E}} p^k \times \frac{\partial c_{n_k}(t-k)}{\partial b} \times \left\{ \xi(t|1) \times \hdots \times \xi(t|k)\right\}, \label{eq:Fbeforelimits}
\end{align}
\end{widetext}
where 
\begin{dmath*}
\xi(t|j)=\left[ (1-p)+p \frac{1}{d(n_{(j-1)})}\sum_{(n_{(j-1)},n_j) \in \mathcal{E}} c_{n_j}(t-j)\right]^{d(n_{(j-1)}) -1}.
\end{dmath*}

In the appropriate limits, Eq. (\ref{eq:Fbeforelimits}) reduces to the condition for the vanishing derivative of $F(b)$, i.e., $0=\partial F/\partial b$. Since $c_n(t)\to b=1$ as $t \to \infty$, we set $\partial c_m(t)/\partial b = 1$ for each $m$. Since the derivatives are being evaluated at $b=1$, in the limit of $t\to\infty$, each $\xi(t|j-1,j)$ evaluates to $1$, and can hence be ignored. We take the limit of $k\to \infty$ to give
\begin{equation*}
0 = 1- \lim_{k\to\infty} \sum_{(n_0,n_1) \in \mathcal{E}} \hdots \sum_{(n_{(k-1)},n_k) \in \mathcal{E}} (p_c)^k,
\end{equation*}
where $n_0$ is the node considered initially, and for each $k$, $n_k$ spans over all nodes reachable after traversing $k$ edges from $n_0$. The summand is now independent of the variables of summation and hence the entire series of summations is reduced to the total number of terms that are summed over. Thus,
\begin{equation*} 
\pcrit = \lim_{k\to\infty} \left(\prod_{l=0}^{k-1} \langle d(n_l)\rangle \right)^{-1/k},
\end{equation*}
where $\langle d(n_l)\rangle$ is the average degree of nodes connected to the node $n_0$ after traversing exactly $l$ edges (under this definition $\langle d(n_0)\rangle=d(n_0)$). We define $\dinf = \lim_{l\to\infty} \langle d(n_l)\rangle$. By our initial assumption of the network being strongly connected, $d(n)>0$ for each node $n$. Thus, $\langle d(n_l)\rangle>0$ for each $l$, and hence we obtain
\begin{equation}\label{eq:pcritonebyd}
\pcrit=\dinf ^{-1}.
\end{equation}

It can be shown that $\dinf$ is the average of the degrees at each node of the network when weighted by the component of $v^{PF}$ at that node, where $v^{PF}$ is the Perron-Frobenius eigenvector of $\widetilde{A}^T$. We use this idea of weighing quantities by $v^{PF}$ later when the assumption of the locally tree-like approximation is broken, since it arises naturally as a consequence of taking limits of $t \to \infty$ in the recursion relation, which in effect includes terms spanning the nodes of the network according to the paths that connect them to $n_0$. We can then show that this weighted average of the degrees, $\dinf$ is just the Perron Frobenius eigenvalue of $\widetilde{A}$, giving the result shown earlier in Eq. (\ref{eq:pcritonebylambda}). The details of the derivations can be found in Appendix \ref{apx:treelikeproof}. 

Thus from Eq. (\ref{eq:pcritonebyd}), this method of determining criticality under the tree-like approximation gives $\pcrit=(\tlambda)^{-1}$ and hence at criticality the Perron-Frobenius eigenvalue for $A$, i.e., $\lambda=\pcrit \times \tlambda=1$, replicating the condition for criticality from Ref.\cite{larremore2012statistical}. 

\subsection{Estimating the critical transmission accounting for corrections due to $2$-parallel motifs}\label{sec:pcritsecondlevel}

In Sec. \ref{sec:secondrecursion} we first setup a recursion relation in analogy to Eq. (\ref{eq:firstlevelrecursion}) that takes into account the effect of $2$-parallel motifs by assuming independence of distributions at times $(t-2)$ (as opposed to independence at times $(t-1)$, as is assumed for the locally tree-like approximation). Then, in Sec. \ref{sec:pcrit}, we use the same idea of evaluating derivatives to find the condition for a double root at $b=1$ to estimate $\pcrit$.

\subsubsection{Recursion relation}\label{sec:secondrecursion}

We rewrite the recursion relation in Eq. (\ref{eq:firstlevelrecursion}) as
\begin{equation}\label{eq:firstlevelalt}
c_{n_0}(t) = \sum_{k=0}^{d(n_0)} \sum_{\mathcal{E}^0_k}(1-p)^{d_{n_0}-k} \; p^k \prod_{v \in \mathcal{V}^0_k}{c_v(t-1)},
\end{equation}
where $\sum_{\mathcal{E}^0_k}$ denotes a sum over all possible sets $\mathcal{E}^0_k$ which are $k$-element subsets of the set of edges that begin at $n_0$; and $\mathcal{V}^0_k$ is the set of nodes to which the edges of the set $\mathcal{E}^0_k$ point.

We can interpret the terms in Eq. \ref{eq:firstlevelalt} as follows: 
the index $k$ counts the number of edges connected to the node $n_0$ that are activated due it and ranges from $0$ to $d(n_0)$; 
the set $\mathcal{E}^0_k$ is the $k$-element set of activated edges; $\mathcal{V}^0_k$ is the corresponding set of activated nodes, which also has $k$ elements, since we have assumed that the network has no double edges;
$p^k$ is the probability that the $k$ edges were activated; 
$(1-p)^{d(n_0)-k}$ is the probability that the remaining $(d(n_0)-k)$ edges remained unactivated; 
and finally, the product term, $$\prod_{v \in \mathcal{V}^0_k}{c_v(t-1)}$$, is the probability that after each of the $k$ activated nodes, all avalanches had a duration of less than or equal to $(t-1)$ units of time. This final product term from Eq. \ref{eq:firstlevelalt} can be rewritten using the same equation as a recursion relation to obtain an expression dependent on probabilities as a function of $(t-2)$. Rather than using the recursion relation to write the expression directly, we rewrite the product term in an equivalent form as
\begin{equation}
\label{eq:product}
\prod_{v \in \mathcal{V}^0_k}{c_v(t-1)} = \sum_{r=0}^{D[\mathcal{V}^0_k]} \sum_{\mathcal{E}^1_r} (1-p)^{D[\mathcal{V}^0_k] - r} \; p^r \prod_{w \in \mathcal{V}^1_r}{c_w(t-2)},
\end{equation}
where we define $D[\mathcal{V}^0_k] = \sum_{v \in \mathcal{V}^0_k} d(v)$ as the sum of the degrees of the nodes in the set $\mathcal{V}^0_k$, which is the total number of edges that begin from the $k$ nodes activated by $n_0$; as earlier, $\sum_{\mathcal{E}^1_r}$ is the sum over all possible sets $\mathcal{E}^1_r$ which are $r$ element subsets of the set of edges, $\mathcal{E}$, that begin anywhere in the set $\mathcal{V}^0_k$; and, $\mathcal{V}^1_r$ is the set of vertices to which the edges of the set $\mathcal{E}^1_r$ point. 

Here, $r$ is the variable that counts the number of edges activated due to any of the $k$ activated nodes at the previous time step, and analogous to the earlier equation, $\mathcal{V}^1_r$ is the set of activated nodes due to the $r$ activated edges. 

In the case of the locally tree-like approximation: all edges present at one edge away from the initial node $n_0$ are independent of each other, and hence the set $\mathcal{V}^1_r$ has exactly $r$ elements; the final product term then represents the probability that the avalanches beginning from these nodes have a duration of no longer than $(t-2)$; and then Eq. (\ref{eq:product}) is equivalent to the original recursion relation in Eq. (\ref{eq:original}).

In the more general case (where the locally tree-like approximation may not be valid), it is possible for some of the edges that are present one edge away from the initial node, $n_0$, to end at the same node. This is due to the presence of $2$-parallel motifs in the network, and in this case the set $\mathcal{V}^1_r$, which contains the nodes activated due to the $r$ activated edges, contains $\leq r$ elements. Thus, we have,
\begin{dmath}\label{eq:secondlevel}
c_{n_0}(t) = \sum_{k=0}^{d(n_0)} \left[ \sum_{\mathcal{E}^0_k}(1-p)^{d(n_0)-k} p^k \\ \times\sum_{r=0}^{D[\mathcal{V}^0_k]}\left( \sum_{\mathcal{E}^1_r} (1-p)^{D[\mathcal{V}^0_k] - r} \; p^r \prod_{w \in \mathcal{V}^1_r}{c_w(t-2)} \right) \right],
\end{dmath}
with $|\mathcal{V}^1_r|\leq r$. We henceforth refer to this equation as the second-level approximation equation.

We now estimate the quantity $|\mathcal{V}^1_r|$, which can then be combined with the the recursion relation of Eq. (\ref{eq:secondlevel} to estimate $\pcrit$ in a similar fashion to the technique used in Sec. \ref{sec:treebased} to calculate $\pcrit$ from Eq. (\ref{eq:firstlevelrecursion}). Since $|\mathcal{V}^1_r|$ is necessarily less than or equal to $r$, we write it as $|\mathcal{V}^1_r|=r-\delta r$.

We define the coefficients $Q_{\alpha}(n_0)$ at a given node $n_0$ to count the number of 2-parallel structures originating at the node $n_0$. In particular, $Q_{\alpha}(n_0)$ is the number of nodes that are reachable from node $n_0$ via exactly $\alpha$ edge-independent paths of length $2$. This is a quantity that is dependent on the topology of the network which can be  measured independent of the dynamics on the network. For a locally tree-like network, $Q_1 = d(n_0)\langle d(n_1)\rangle$, where as earlier $\langle d(n_1)\rangle$ is the average degree of nodes connected to the node $n_0$ after traversing exactly one edge, and $Q_{\alpha}=0$ for $\alpha>1$. 
In terms of these coefficients, we show in Appendix \ref{apx:deltar}, that by accounting for 2-parallel motifs we can approximate $\delta r$ as
\begin{dmath}\label{eq:deltarapproximation0}
\delta r = \sum_{\alpha} Q_{\alpha}(n_0) \left( \frac{D[\mathcal{V}^0_k]}{d(n_0) \langle d(n_1) \rangle}\right)^{\alpha} \\ \times \left[ \alpha \left( \frac{r}{D[\mathcal{V}^0_k]}\right) - 1 + \left( 1- \frac{r}{D[\mathcal{V}^0_k]}\right)^\alpha \right]. \label{eq:deltar}
\end{dmath}

Equations (\ref{eq:secondlevel}) and (\ref{eq:deltar}) contain all the required information to treat the second-level approximation equation in a similar fashion to Eq. (\ref{eq:firstlevelrecursion}) and derive the conditions and equations for $\pcrit$.

While the sum over $\alpha$ in principle goes up to a maximum value of $\alpha = D[\mathcal{V}^0_k]$, for further simplification we can apply a cut-off on $\alpha$, by only considering terms for which $\alpha<\alpha_{thr}$.
 We apply this cut-off on both the $Q_{\alpha}(n_0)$ coefficients, as well as in the expansion of $(1-r/D[\mathcal{V}^0_k])^{\alpha}$. The cutoff is justified because when averaging across nodes $Q_{\alpha}(n_0)$ falls off very rapidly with $\alpha$, i.e., there are few nodes with significant values of $Q_\alpha(n_0)$ for large $\alpha$, while most nodes only have nonzero values of $Q_\alpha(n_0)$ for small values of $\alpha$. 
Further, $r/D[\mathcal{V}^0_k]$ can be assumed to be much smaller than $1$, since terms for larger $r$ are exponentially suppressed with a $p^r$ term in the second level recursion relation, and would correspond to a large fraction of the edges connected from a single node being activated simultaneously in the same time step. Hence we can apply a cut-off on the expansion of $(1- r/D[\mathcal{V}^0_k])^\alpha$ as well. 

Note that $Q_\alpha(n_0)$ is the number of nodes that are reachable from $n_0$ via exactly $\alpha$ edge-independent paths of length two. Since our algorithm for network generation (described in Sec. \ref{sec:numerics}) involves adding new paths of length two between nodes that already are separated by a path of length two, many of our generated networks tend to have $Q_\alpha>0$ for larger values of $\alpha$ than might be normally expected in real-world networks. We find that for the networks that we have constructed, and for other networks we have observed, $Q_\alpha\approx0$ for $\alpha>6$, and hence for our final results we use a cutoff of $\alpha_{thr}=6$.
For the remainder of Sec. \ref{sec:twoparallel} however, we use the an approximation to $\delta r$ assuming a cut-off of $\alpha_{thr}=3$ as a representative cut-off to demonstrate the subsequent algebra. This gives us
\begin{dmath}
\delta r = \left\{ \frac{Q_2(n_0)}{[d(n_0)\langle d(n_1)\rangle]^2} + \frac{3 Q_3(n_0)}{[d(n_0)\langle d(n_1)\rangle]^3} D[\mathcal{V}^0_k] \right\}r^2 - \frac{Q_3(n_0)}{[d(n_0)\langle d(n_1)\rangle]^3}r^3 . \label{eq:deltarapproximation}
\end{dmath}
It should be noted that this cut-off can be made higher without any significant change to the method of analysis presented below. In general, for a cut-off at $\alpha_{thr}$, the resulting approximation for $\delta r$ is an $(\alpha_{thr} - 2)$ degree polynomial in $D[\mathcal{V}^0_k]$. We present relevant expressions of our final results for larger values of the cut-off in Appendix \ref{apx:cutoffexpressions}.

\subsubsection{Finding the critical transmission probability}\label{sec:pcrit}

Treating the addition of $2$-parallel motifs to the network structure as a small change to the overall network dynamics around the new point of criticality in the network, we use the same method for estimating $\pcrit$ as discussed previously in Sec. \ref{sec:treebased}. The derivatives are evaluated of the second-level approximation equation, Eq. (\ref{eq:secondlevel}), after which appropriate limits are taken. 

\begin{widetext}
\begin{equation*}
\frac{\partial c_{n_0}(t)}{\partial b} = \sum_{k=0}^{d(n_0)} \left\{ \sum_{\mathcal{E}^0_k}(1-p)^{d(n_0)-k} p^k \sum_{r=0}^{D[\mathcal{V}^0_k]}\left[ \sum_{\mathcal{E}^1_r} (1-p)^{D[\mathcal{V}^0_k] - r} \; p^r \left(\left\{ \prod_{w \in \mathcal{V}^1_r}{c_w(t-2)} \right\}  \sum_{w \in \mathcal{V}^1_r} \frac{1}{c_w(t-2)}\frac{\partial c_w(t-2)}{\partial b}\right)\right] \right\}. 
\end{equation*}
\end{widetext}

Once again, since the derivatives are evaluated at $b=1$, several terms evaluate to $1$ in the appropriate limits as performed earlier in Sec. \ref{sec:treebased}, yielding
\begin{dmath}\label{eq:secondlevelrecursion}
1= \sum_{k=0}^{d(n_0)} \left\{ \sum_{\mathcal{E}^0_k}(1-p)^{d(n_0)-k} \; p^k \times \\ \sum_{r=0}^{D[\mathcal{V}^0_k]}\left[ \binom{D[\mathcal{V}^0_k]}{r} (1-p)^{D[\mathcal{V}^0_k] - r} p^r \left( r - \delta r \right)\right] \right\},  
\end{dmath}
which results from using $|\mathcal{V}^1_r|=r-\delta r$, and observing that $\mathcal{E}^1_r$ is an $r$ element subset of a $D[\mathcal{V}^0_k]$ element superset of edges. Thus, there are $\binom{D[\mathcal{V}^0_k]}{r}$  such subsets, and the summation over $\mathcal{E}^1_r$ contains $\binom{D[\mathcal{V}^0_k]}{r}$ terms in the summation. 
We can now use the previously derived approximation for $\delta r$, Eq. (\ref{eq:deltarapproximation}), to simplify the above expression. 

The term in the square brackets when summed over $r$ from $0$ to $D[\mathcal{V}^0_k]$ is equivalent to the expectation value of the $3^{\text{rd}}$ degree polynomial in $r$, i.e., $r-\delta r$ (with $\delta r$ given by Eq. (\ref{eq:deltarapproximation})) over a binomial distribution with probability $p$ over $D[\mathcal{V}^0_k]$ trials. This simplifies to give a $3^{\text{rd}}$ degree polynomial in $D[\mathcal{V}^0_k]$. Further, since the only term dependent on the set $\mathcal{E}^0_k$ is $D[\mathcal{V}^0_k]$, to simplify the sum over all such sets it suffices to evaluate this sum for powers of $D[\mathcal{V}^0_k]$. We demonstrate how to evaluate these sums in Appendix \ref{apx:sumDSk}. 
Using this, we simplify the final binomial summation over $k$ to obtain a polynomial equation for $p$ as 
\begin{dmath}\label{eq:secondlevelpcritsingle}
1=\left(\frac{p^2}{d(n_0)^2 \langle d(n_1) \rangle^3}\right)\times \left\{ d(n_0)^3 \langle d(n_1) \rangle^4 - d(n_0) \langle d(n_1) \rangle Q_2(n_0) \left[ \langle d(n_1) \rangle - p \langle d(n_1) \rangle + p \langle d(n_1)^2 \rangle + p^2 \langle d(n_1) \rangle^2(d(n_0) - 1)\right] + Q_3(n_0) \left[ -3 \langle d(n_1)^2 \rangle (p-1)^2 + p \langle d(n_1)^3 \rangle (p-3) -3 p (p-1)^2 \langle d(n_1) \rangle^2 (d(n_0) -1) + p^3 (p-3) \langle d(n_1) \rangle^3 (d(n_0) -1) (d(n_0) -2) + \langle d(n_1) \rangle %
(1 -3p +  2p^2+3p^2 (p-3) (d(n_0)-1) \langle d(n_1)^2 \rangle)\right]\right\}.
\end{dmath}

Under the locally tree-like approximation, if we evaluate derivatives for only a single step in the recursion relation of Eq. (\ref{eq:firstlevelrecursion}), we obtain $1=p \times d(n_0)$. Then, in the process of taking the appropriate limits, this form can be reduced to $1 = p_c \times \dinf$. We analogously posit that if we look at the right-hand side of Eq. (\ref{eq:secondlevelpcritsingle}) and evaluate each term not at $n_0$ or $n_1$, but rather in terms of an average over $v^{PF}$ as earlier, then the resulting equation allows us to determine $\pcrit$ by taking appropriate limits in Eq. (\ref{eq:secondlevelrecursion}). 
This gives the following equation to be solved for $\pcrit$ 

\begin{dmath}\label{eq:secondlevelpcrit}
1=\left(\frac{\pcrit^2}{\dinf^5}\right)\times \left\{ \dinf^7 - \dinf^2 \overline{Q_2} \left[ \dinf - \pcrit \dinf + \pcrit \langle d^2 \rangle + \pcrit^2 \dinf^2(\dinf - 1)\right] + \overline{Q_2} \left[ -3 \langle d^2 \rangle (\pcrit-1)^2 + \pcrit \langle d^3 \rangle (\pcrit-3) -3 \pcrit (\pcrit-1)^2 \dinf^2 (\dinf -1) + \pcrit^3 (\pcrit-3) \dinf^3 (\dinf -1) (\dinf -2) %
+ \dinf (1 -3\pcrit + 2\pcrit^2+3\pcrit^2 (\pcrit-3) (\dinf-1) \langle d^2 \rangle)\right]\right\},
\end{dmath}
 where $\langle d^q \rangle$ is the average of the $q^{\text{th}}$ power of the degrees in the network when weighted by $v^{PF}$, and $\overline{Q_q}$ is the quantity $Q_q(n_0)$ averaged over all nodes, weighted by $v^{PF}$. For the case of a network satisfying the locally tree-like approximation, i.e., when there are no $2$-parallel motifs in the network, $\overline{Q_q}=0$ for all $q>1$, and hence the above equation reduces to $\pcrit=1/\dinf$, which is the same result as obtained in Eq. (\ref{eq:pcritonebyd}). Similar to the expression derived for the case of $\alpha_{thr}=3$, we can derive polynomial expressions in $\pcrit$ for higher values of $\alpha_{thr}$. For the results presented in Sec. \ref{sec:numerics} we have used a cut-off of $\alpha_{thr}=6$. The polynomial expression solved in this case has been shown in Appendix \ref{apx:cutoffexpressions}. Solving these high-degree polynomial expressions generally results in multiple roots for $\pcrit$. However, we find numerically that only one root satisfies $0<\pcrit<1$, which is the solution that we use. 

We can also use the coefficients $\overline{Q_{\alpha}}$ to construct a single network parameter that we can use to compare the number of $2$-parallel motifs in different networks. In Eq. (\ref{eq:deltarapproximation0}) if we set $r$ to its maximum possible value of $D[\mathcal{V}^0_k]$, and set $D[\mathcal{V}^0_k]$ to its maximum possible value of $d(n_0) \langle d(n_1) \rangle$, then $\delta r = \sum_{\alpha} [Q_{\alpha}(n_0)\times (\alpha-1)]$. For this $\delta r$, the quantity $r-\delta r$ represents the total number of unique nodes that can be reached by traversing two edges. To normalize $\delta r$, we consider the ratio $\delta r/r = \sum_{\alpha} [Q_{\alpha}(n_0)\times (\alpha-1)] / d(n_0) \langle d(n_1) \rangle$. Accordingly, we define the parameter $\sigma$, which we use in Sec. \ref{sec:numerics} to compare the networks of different sizes, as
\begin{equation}\label{eq:sigma}
\sigma = \frac{\sum_{\alpha} [\overline{Q_{\alpha}}\times (\alpha -1)]}{\dinf^2}.
\end{equation}
Note that $0\leq\sigma\leq 1$, and for a network with all-to-all connections, $\sigma=1$.

We now go back to the earlier claim that for graphs with large average degree, a tree-based theory is valid, independent of the number of $2$-parallel motifs in the network (see Fig. \ref{fig:varyingd}). Let us assume that the nodes in $\mathcal{V}^0_k$ have some large average degree $d \gg 1$. Let us consider the case of the highest density of 2-parallel motifs where we assume that each of the nodes in $\mathcal{V}^0_k$ connect to the same $d$ nodes. If we consider the set $\mathcal{V}^1_r$ (set of vertices reached from an $r$ element subset of edges starting at  $\mathcal{V}^0_k$), the probability of a certain value of the cardinality $|\mathcal{V}^1_r|$ over the ensemble of all possible activated $r$ edges is given by
\begin{equation*}
\Pr(|\mathcal{V}^1_r|) = \left. \binom{d}{|\mathcal{V}^1_r|} \times \binom{r-1}{|\mathcal{V}^1_r|} \middle/ \binom{r+d-1}{d-1}\right. ,
\end{equation*}
which is obtained by considering the number of ways to first choose the $|\mathcal{V}^1_r|$ out of the $d$ vertices to get activated, and then count the number of ways to be able to write $r$ as the sum of $|\mathcal{V}^1_r|$ positive integers. We then divide by the total number of possibilities, which are the number of ways to write $r$ as the sum of $d$ non-negative integers
Under this distribution, the expected value of $|\mathcal{V}^1_r|$ is $r d /(r+d-1)$, which in the limit $d\to\infty$ gives $|\mathcal{V}^1_r|\to r$. Thus $\delta r \to 0$ and hence networks with large average degree can be treated directly under the locally tree-like approximation. 

We further examine this heuristic argument in the context of Eq. (\ref{eq:secondlevelpcrit}). Since $\tlambda=\dinf$ (as derived in Appendix \ref{apx:treelikeproof}) we write $\lcrit=\pcrit\tlambda=\pcrit \dinf$ and note that $\lcrit\sim\scO(1)$ to imply that $\pcrit\sim \scO(\dinf^{-1})$. Thus, if $\scO(\overline{Q_\alpha})<\scO(\overline{Q_2})$ for $\alpha>2$, we see from Eq. (\ref{eq:secondlevelpcrit}) that
\begin{equation}\label{eq:degreeheuristicnoapprox}
\lcrit = 1+ \overline{Q_2}\left[\scO\left( \frac{1}{\dinf^4}\right) +\scO\left( \frac{\langle d^2\rangle}{\dinf^6}\right) -\scO\left( \frac{\langle d^3 \rangle}{\dinf^8}\right) \right].
\end{equation}
We first consider the above equation under the approximation of a tightly-peaked degree distributions, such that $\langle d^q \rangle \sim \scO(\dinf^q)$. The expression in the square brackets is dominated by the leading term of $1/\dinf^4$, and hence
\begin{equation}\label{eq:degreeheuristic}
\lcrit = 1+ \overline{Q_2}\scO\left( \frac{1}{\dinf^4}\right).
\end{equation}
Thus in the limit of large degrees $\lcrit=1$ and $\pcrit=1/\dinf = (\tlambda)^{-1}$. However, when the fourth power of the average degree is large compared to 1, deviations from a tree-based theory are suppressed.

What happens in the case that networks do not satisfy the an approximation of a tightly-peaked degree distribution? In this case, where $\langle d^q \rangle \sim \scO(\dinf^q)$ is no longer valid, the magnitude of the terms in the terms in the square brackets cannot be compared in a straight-forward fashion and it is unclear which of the positive or negative terms dominates over the others. In fact, for the case of scale-free degree distributions, averages over higher powers of degrees may diverge and not be well defined in the limit of infinitely large networks. However, as verified in Fig. \ref{fig:swaps}(b), for large-but-finite network sizes our derived result Eq. (\ref{eq:secondlevelpcrit}) continues to hold, and only a small deviations from a tree-based theory are observed for a network with average degree as low as $3.6$. Further, as can be seen in Fig. \ref{fig:varyingd}, deviations away from a tree-based theory decay at a similar rate to the case for sharply peaked distributions. Thus despite a straight-forward comparison for increasing average degree not being possible for scale-free networks from Eq. (\ref{eq:degreeheuristicnoapprox}), numerical experiments on large-but-finite networks suggest a similarly rapid decay of $\lcrit$ towards 1.

In Fig. \ref{fig:swaps} the deviation away from a tree-based theory for a scale-free degree distribution is marginally greater than the deviation in the case of a Poisson degree distribution when compared at the same $\sigma$ with despite having a slightly larger average degree (3.6 for the scale-free network versus 3.3 for the poissonian network). See Fig. \ref{fig:comp3} for an explicit comparison. 
As noted above however, this is not a systematic comparison apparent from Eq. (\ref{eq:degreeheuristicnoapprox}); In Appendix \ref{apx:algointermediate} we present an example of a network with a degree distribution that is intermediate between a scale-free degree distribution and a tightly-peaked degree distribution which exhibits a larger deviation at the same $\sigma$ and average degree. Thus our results do not give a clear interpretation to the expected magnitude of deviation away from a tree-based theory dependent on the nature of the degree distribution. However, in each case we observe from Eq. (\ref{eq:secondlevelpcrit}) that the deviations from a tree-based theory appear to be of small magnitudes even at low average degree, and numerical results in each case demonstrate that our theory continues to give good results in all cases.

In practice (as demonstrated in Fig. \ref{fig:varyingd}), we see that for large networks with average degrees $\gtrsim5$, often give deviations of less than $0.1\%$ between the empirically determined $\lcrit$, and $\lcrit=1$ as predicted by the tree-based theory. Since most networks that are encountered in physical systems tend to have large average degrees (for example almost all networks considered by Melnik et. al\cite{melnik2011unreasonable}), the locally tree-like approximation is usually a useful tool.



\section{Discussions and Conclusions}\label{sec:conclusion}

In studying criticality of network cascades, tree-based approximations have been shown to often be accurate for a wide variety of real world networks\cite{melnik2011unreasonable}. We develop analytic reasoning to justify the effectiveness of this approximation in predicting criticality for network dynamics that are a variant of discrete time SIS epidemiological models on a network, i.e., dynamics that allow for nodes to be re-excited an arbitrary number of times (model description in Sec.\ref{sec:branchingprocess}).

Under the locally tree-like approximation, the condition for criticality of network avalanches is that the largest eigenvalue $\lambda$ of the probability weight matrix is one.  Examining the assumptions made in deriving conditions of criticality according to a tree-based theory, we study the factors that contribute to deviations introduced when the locally tree-like approximation is no longer valid (Sec.\ref{sec:failure}). We show that the network motifs contributing to these deviations are the $k$-parallel motifs (see Fig. \ref{fig:kparallel}), the smallest of which is the $2$-parallel motif (also known as the bi-parallel motif). In particular, for the simple discrete time SIS model that we study, the presence of a high density of triangular motifs does not imply large deviations away from a tree-based theory. In addition, large densities of 2-parallel motifs do not necessarily imply large deviations from a tree-based theory --- large average degree can suppress these deviations independent of the density of 2-parallel motifs. For example, critical points for large all-to-all networks are well predicted by tree-based theories. We also explain how tree-based theories necessarily under-predict the required largest eigenvalue for criticality, i.e., $\lcrit\geq 1$.

To study critical network cascades in networks that may not satisfy the locally tree-like approximation, we derive an expression for $\lcrit$ on network topologies that allow for the presence of 2-parallel motifs (Sec.\ref{sec:derivations}). 
We obtain a polynomial equation for the transmission probability, $p_c$, corresponding to criticality in the network, which depends on the density of 2-parallel motifs, $\sigma$, in the network. Given $p_c$ we get $\lcrit$ from the direct relationship: $\lcrit=p_c \tilde{\lambda}$, where $\tilde{\lambda}$ is the Perron-Frobenius eigenvalue of the network's adjacency matrix $\tilde{A}$. We use the obtained expressions to demonstrate that for networks with large average degree, the relevance of the $2$-parallel motifs is reduced, and results according to a locally tree-like approximation agree well with empirically determined quantities.

We then verify the derived equation, by performing numerical simulations of the described dynamics on a class of networks generated to have a large number of $2$-parallel motifs, and hence exhibit a critical transmission probability that is different as compared with the predictions of the locally tree-like approximation. We observe that the predictions made by taking into account the effect of $2$-parallel motifs agree well with empirical observations made on these generated networks, demonstrating the effectiveness of our derived results (Sec. \ref{sec:numerics}, Fig. \ref{fig:swaps}). 

For networks with tightly-peaked degree distributions we show that deviations from a tree-based theory will be suppressed by a factor of the fourth power of the degree (see Eq. (\ref{eq:degreeheuristic})). For networks with skewed degree distributions, although a straight-forward relationship with average degree is not apparent, application of our results to generated scale-free networks continues to show a similarly rapid decay of deviations from a tree-based theory with increasing degree as can be seen in Fig. \ref{fig:varyingd}. 
Thus, we note that when the average degree is even modestly large -- more specifically, when the fourth power of the average degree is large compared to 1 -- deviations away from a tree-based theory are strongly suppressed. Empirically, from Fig. \ref{fig:varyingd}, we observe that an average degree as small as 10 is sufficient to almost entirely supress any deviations. Since most real-world networks (including a large majority of networks considered in Ref. \cite{melnik2011unreasonable}) have large average degrees that are at least modestly large, we expect tree-based theories to provide sufficiently accurate descriptions of the network dynamics.

It is important to note that we have restricted our analysis to the limit of large network sizes. In general, small networks also promote the interaction between network cascades and contribute to deviations away from a tree-based theory. In particular, small networks also promote larger than expected values of $\lcrit$. We expect that in general with increasing network sizes the predictions made via our `second-level approximation' will agree with empirical estimates for criticality (as measured via the $\kappa$ metric\cite{shew2009neuronal}) with increasing accuracy. For skewed degree-distributions, in the limit of an infinite network size, the averages over higher powers of degrees may diverge. 
However, for large but finite network sizes we expect our derived expressions to be valid, and our broader conclusions of suppression of deviations from a tree-based theory with increasing average degree to continue to hold. 

We note that our analysis assumes the system is near criticality. While empirical evidence suggests that tree-based approximations appear to hold away from criticality\cite{melnik2011unreasonable}, our analysis does not address this case.

Note that we assume the independence of distributions at two time steps before the current time, as opposed to independence at a single time step before the current time as is done in a tree-based theory. This idea may in principle be extended towards a greater number of time steps to observe higher-order effects relating to the deviations from a tree-based theory. 

While we have focused on the case of SIS-like dynamics on networks (i.e., where nodes are re-excitable), we note that in the case of a directed network that \emph{does} satisfy the locally tree-like approximation, SIS-like dynamics are identical to SIR-like dynamics (i.e., percolation problems, wherein after activation, nodes are removed from the network) near the critical point, since if excitation paths of short length do not interact, then the re-excitability of a node is immaterial to the system dynamics. In our analysis of SIS dynamics corresponding to Eq. (\ref{eq:dynamics}), we argued that the 2-parallel motif was the smallest motif relevant to deviations away from a tree-based theory, and studied the correction introduced to Eq. (\ref{eq:product}) as a result of 2-parallel motifs in the network [Eqs. (\ref{eq:secondlevel}, \ref{eq:deltarapproximation0})]. Analogously, we expect these ideas to apply to SIR-like dynamics as well, by including the correction to Eq. (\ref{eq:product}) as a result of feed-forward triangular motifs, which is the smallest relevant motif in this case (in a discrete time SIR model, if a node $x$ activates nodes $y$ and $z$, and node $y$ then attempts to activate $z$ again, $z$ will remain inactive, having been effectively removed from the network following the first time it was activated. Thus such motifs result in behavior that is not locally tree-like, and would manifest as fewer activated nodes as compared to a locally tree-like network). Based on the number of triangular motifs and 2-parallel motifs, SIR-like dynamics result in a different expression for $\delta r$ in Eq. (\ref{eq:deltarapproximation0}), which can be used in the remainder of our analysis to construct an analogous expression for $\lcrit$. We leave the details of this study for future research work. 

We expect the results for $\lcrit$ for SIR-like dynamics to similarly demonstrate that in the limit of large average degree, tree-based theories will result in close agreement to observed dynamics on networks that are far from trees. While we do not show this analysis here, we now consider a motivating example of a large-$N$, all-to-all network, and demonstrate that networks with a high density of triangular motifs will continue to behave tree-like in the limit of large average degrees.
Starting from any node in the network, consider the ensemble of pairs of paths of length 2 starting at that node. The total number of such pairs of paths scales as $N^4$. We compare this with the number of pairs of paths that result in dynamics that are not locally tree-like, i.e., pairs of paths that meet each other at any point. There are $N^3$ pairs of paths that meet each other at the end points, corresponding to deviations as a result of 2-parallel motifs. This is relevant to both SIS-like dynamics (as discussed earlier in Sec. \ref{sec:failure}) and SIR-like dynamics. However, for SIR-like dynamics, we also need to consider contributions due to feed-forward triangular motifs formed by these pairs of paths. Again, there are $N^3$ pairs of paths that result in such a motif. Since we are considering an all-to-all network, each of the pairs of paths will be equally likely, and thus the probability that excitation paths will interact, leading to dynamics that are not locally tree-like scales as $N^3/N^4 \sim 1/N$, which tends to zero as $N$ goes to $\infty$. Thus, in a large all-to-all network the probability that the network will display deviations away from a tree-based theory is negligible for both SIS-like and SIR-like dynamics.

Thus, we expect the type of analysis we have developed here to not only apply to SIS-like dynamics, but also to SIR-like dynamics, and hence also to percolation problems on complex networks. Consequently, we expect our ideas and results to provide insight towards a wide range of problems in network science relating to critical dynamics.

\section{Acknowledgments}
This work was supported by the National Science Foundation through Award \# 1632976 and by the Army Research Office under Grant No. W911NF-12-1-0101. 
\appendix

\section{Effects of network size}\label{apx:bigenough}

\begin{figure*}
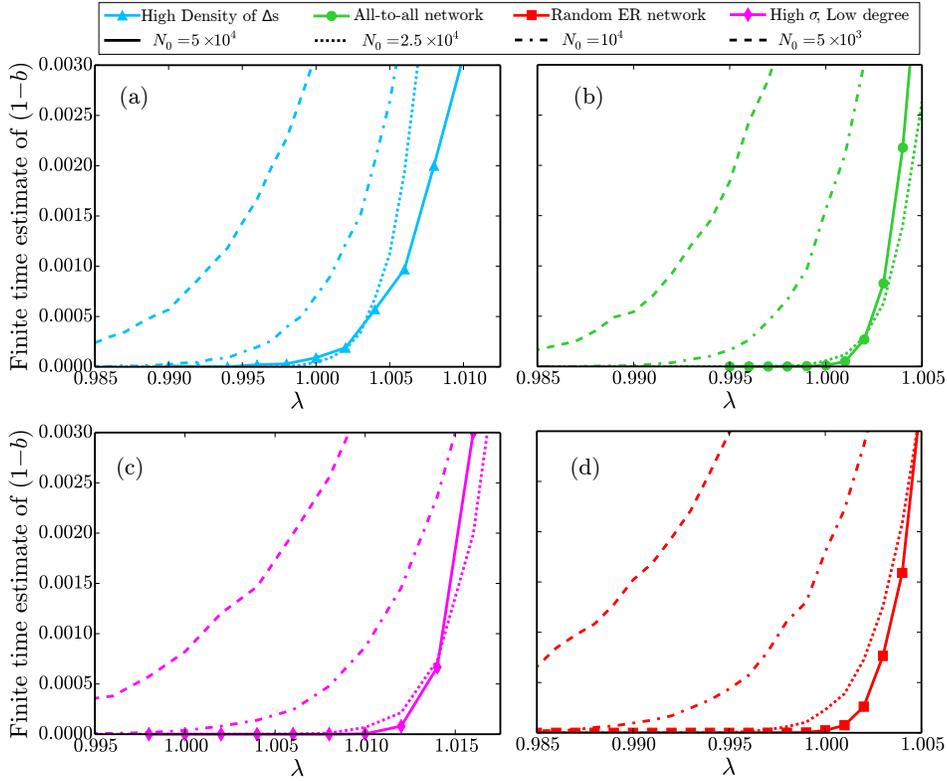

\includegraphics[width=1.5\columnwidth]{{{Phase_transitions_all_curves_vary_N_new}}}
\caption{Phase transition curves for networks with varying network sizes, the four types of networks considered in Fig. \ref{fig:phasetransition}. In each case, the solid curve with markers represents the largest network size with $N_0=5\times10^4$ (corresponding to the size used in text of the paper), the dotted curve represents $N_0=2.5\times 10^4$, the dash-dotted curve represents $N_0=10^4$ and the dashed curve represents $N_0=5\times 10^3$. Since the curves begin to approach each other at the largest network sizes considered, we use $N_0=5\times 10^4$ as finite-size approximation to the infinite network size limit. See Appendix \ref{apx:bigenough} for details on how these networks were generated, and the caption of Fig. \ref{fig:phasetransition} for the interpretation of such phase transition curves in the context of criticality.
}
\label{fig:combinedphasetransition}
\end{figure*}

In our paper we generally perform our numerical simulations on networks with $N\lesssim N_0=5\times 10^5$ nodes. To demonstrate that this network size is sufficiently large, we present a series of phase transition curves similar to Fig. \ref{fig:phasetransition} for varying network sizes. In Fig. \ref{fig:combinedphasetransition} we show a series of 4 curves for each type of network considered: 
\begin{enumerate}[(a)]
	\item Networks with a high density of feed-forward triangular motifs, but a negligible density of $k$-parallel motifs. These networks were generated by initially considering $N_0$ nodes, then considering $N_0$ random triples which are then connected in a feed-forward fashion., and finally by taking the strongly connected component of the resulting network. The network obtained will have $N$ nodes for $N\lesssim N_0$
	\item All-to-all networks on $N_0$ nodes. 
	\item Networks with a high density of $2$-parallel motifs, generated using the algorithm shown in Appendix \ref{apx:algoER}
	\item The strongly connected component of Erd\H{o}s-R\'{e}nyi networks corresponding to the networks in panel (c).
\end{enumerate}

In each case, $N_0$ was varied from $5\times 10^3$ to $5\times 10^4$. Since only the strongly connected component of the network was always considered, the final network sizes, $N$, are slightly smaller than $N_0$ (except for the case of all-to-all networks, where $N=N_0$). The solid curves with markers represent this largest network size, and have also been shown in Fig. \ref{fig:phasetransition}. Since it appears that increasing network sizes results in phase transition curves approaching the curve corresponding to $N_0=5\times 10^4$, we use this largest size for all of our numerical simulations.

\section{Definition of the $\kappa$ metric}\label{apx:kappadefinition}

The $\kappa$ metric was first introduced in Ref. \cite{shew2009neuronal} in the context of cascades of neuronal excitation in the cortex. In this paper we use the python package \texttt{powerlaw} written by Alstott et al\cite{alstott2014powerlaw}, which implements the following definition of $\kappa$
\begin{equation}
\kappa = 1 + \langle F^{fit}(s) - F^{obs}(s) \rangle_s,
\end{equation}
where $F^{fit}$ is the cumulative distribution of the best-fit power-law distribution through the data (fit using the techniques of Refs. \cite{clauset2009power,klaus2011statistical}), $F^{obs}$ is the cumulative distribution of the observed data, and the $\langle \cdots \rangle_s$ denotes an average over logarithmically spaced bins over the data. This results in a sensitive metric for measuring deviations away from critical avalanches\cite{shew2009neuronal}, and is such that $\kappa<1$, $\kappa=1$ and $\kappa>1$ corresponds to subcritical, critical and supercritical network avalanches.

\section{Network construction algorithms to generate high densities of 2-parallel motifs}\label{apx:algorithms}

To construct networks with high densities of 2-parallel motifs, we devise algorithms that connect random nodes to triplets of connected nodes in a fashion that generates a 2-parallel motif. The precise algorithm used to make these random choice governs the degree distribution of the resultant network. Below we present three algorithms to generate networks with high density of 2-parallel motifs, the first having a Poisson degree distribution, the second having a scale-free degree distribution, and the third having a degree distribution intermediate between the two. The out-degree distributions for networks generated following each of the three algorithms are shown in Fig. \ref{fig:comp3}(a). A comparison between the observed deviations from the locally-tree like approximation in each of the three networks is shown in Fig. \ref{fig:comp3}(b).

For use in each of the algorithms below, we define the set $triplets$ such that $(x,y,z)\in triplets$ if and only if there is a directed edge from $x$ to $y$ and from $y$ to $z$, and the set $tnodes$ such that $x \in tnodes$ if and only if $(x,y,z)\in triplets$ for some node $y$ and some node $z$.

\subsection{Poisson degree distribution}\label{apx:algoER}

\begin{itemize}
	\item Consider a set of $N_0$ nodes with no connections made initially between the nodes. Since there are no edges in the network yet, we initialize the sets $triplets$ and $tnodes$ as two empty sets.
	\item Randomly choose $M_1 \ll N_0^2$ pairs of nodes and connect them via directed edges to generate an initial seed network. As each edge is connected update the set $triplets$ and $nodes$
	\item For $M_2$ iterations do the following:
	\begin{itemize}
		\item Choose a random node $w$ uniformly from the set of $N_0$ nodes
		\item Choose a random node $x$ uniformly from the set $tnodes$. For this node $x$, choose a random triplet $(x,y,z)$ from $triplets$ such that the first element of the triplet is $x$.
		\item Use the four nodes $x$, $y$, $z$ and $w$ to make a 2-parallel motif by making directed edges from $x$ to $w$ and from $w$ to $z$
		\item Update the set $triplets$ and then the set $tnodes$
	\end{itemize}
	\item	Finally, to ensure that the networks that we use to test our predictions are strongly connected, we then take the strongly-connected component of the network generated. 
\end{itemize}

This results in a network with a high density of 2-parallel motifs with a degree distribution that is approximately Poisson. Unless otherwise specified, we choose $N_0=5\times 10^4$, $M_1 = N_0$, and $M_2=N_0$. To vary the average degree of this network the parameter $M_2$ is varied.

\subsection{Scale-free degree distribution}\label{apx:algoscalefree}

\begin{itemize}
	\item Consider a set of $N_0$ nodes with no connections made initially between the nodes. Since there are no edges in the network yet, we initialize the sets $triplets$ and $tnodes$ as two empty sets.
	\item Consider the nodes $1,2,\hdots,M_1 \ll N_0^2$. Choose $M_2$ pairs of nodes among these $M_1$ nodes and connect them via directed edges to generate an initial seed network. As each edge is connected update the set $triplets$.
	\item For each node $w$ among the remainder of the nodes $(M_1+1),(M_1+2),\hdots,N_0$ do the following $m$ times:
	\begin{itemize}
		\item Choose a random triplet $(x,y,z)$ uniformly from the set $triplets$. 
		\item Use the four nodes $x$, $y$, $z$ and $w$ to make a 2-parallel motif by making directed edges from $x$ to $w$ and from $w$ to $z$
		\item Update the set $triplets$
	\end{itemize}
	\item	Finally, to ensure that the networks that we use to test our predictions are strongly connected, we then take the strongly-connected component of the network generated. 
\end{itemize}

Since choosing the triplet uniformly from the set of triplets inherently biases the choice of nodes $x$ and $z$ to be proportional to their out-degree and in-degree respectively, thus the edges are created in a preferential attachment fashion, similar to the Barab\'{a}si-Albert model used to generate scale-free networks. This algorithm results in a network with a high density of 2-parallel motifs with a degree distribution that is approximately scale free. Unless otherwise specified, we choose $N_0=5\times 10^4$, $M_1 = N_0/5$, $M_2=2M_1$, and $m=2$. To vary the average degree of this network the parameter $m$ is varied.

\subsection{Intermediate degree distribution}\label{apx:algointermediate}

\begin{itemize}
	\item Consider a set of $N_0$ nodes with no connections made initially between the nodes. Since there are no edges in the network yet, we initialize the sets $triplets$ and $tnodes$ as two empty sets.
	\item Randomly choose $M_1 \ll N_0^2$ pairs of nodes and connect them via directed edges to generate an initial seed network. As each edge is connected update the set $triplets$ and $nodes$
	\item For $M_2$ iterations do the following with a $95\%$ probability:
	\begin{itemize}
		\item Choose a random node $w$ uniformly from the set of $N_0$ nodes
		\item Choose a random triplet $(x,y,z)$ uniformly from the set $triplets$. 
		\item Use the four nodes $x$, $y$, $z$ and $w$ to make a 2-parallel motif by making directed edges from $x$ to $w$ and from $w$ to $z$
		\item Update the set $triplets$
	\end{itemize}
	\item and with $5\%$ probability
	\begin{itemize}
		\item Choose 4 random nodes $x,y,z$ and $w$ uniformly from the set of $N_0$ nodes
		\item Use the four nodes $x$, $y$, $z$ and $w$ to make a 2-parallel motif by making directed edges from $x$ to $y$, $x$ to $w$, $y$ to $z$, and from $w$ to $z$
		\item Update the set $triplets$
	\end{itemize}	
	\item	Finally, to ensure that the networks that we use to test our predictions are strongly connected, we then take the strongly-connected component of the network generated. 
\end{itemize}

\begin{figure*}
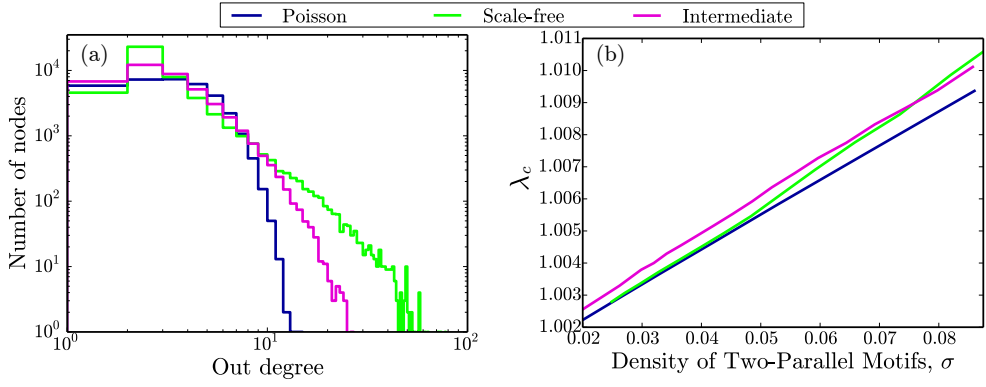

\includegraphics[width=1.5\columnwidth]{{{Comparison3}}}
\caption{(a): Comparison of the out-degree distribution for the networks generated following the algorithms presented in Appendix \ref{apx:algorithms}; (b): Comparison of predicted value of $\lcrit$ according to our derived second-level approximation equations. As noted in the main text, there is no clear trend based on the degree distribution which indicates the expected magnitude of deviation away from the locally-tree like approximation (which predicts $\lcrit=1$ at all $\sigma$). Each of the three curves is approximately linear, appears to intersect $\lcrit=1$ at $\sigma=0$ as would be expected. In both figures, the network generated following Appendix \ref{apx:algoER}  is shown in dark blue	(dark gray), Appendix \ref{apx:algoscalefree} is shown in green (light gray), and Appendix \ref{apx:algointermediate} is shown in pink (an intermediate shade of gray). The blue and red curves in (b) are identical to the solid black curves shown in Fig. \ref{fig:swaps}(a) and \ref{fig:swaps}(b) respectively.
}
\label{fig:comp3}
\end{figure*}

This results in a network with a high density of 2-parallel motifs with a degree distribution that is approximately intermediate between the sharply peaked Poisson distribution and the scale-free distribution for the above two algorithms (see Fig. \ref{fig:comp3}(a) ). We choose $N_0=5\times 10^4$, $M_1 = N_0$, and $M_2=N_0$.

\section{Proof that $\dinf=\tlambda$}\label{apx:treelikeproof}

Let $v^{PF}$ be the Perron-Frobenius eigenvector of $\widetilde{A}^T$, and let $\tlambda$ be the corresponding eigenvalue.
We first show that $\dinf$ is the average of the degrees at each node of the network when weighted by the component of $v^{PF}$ at that node. Then we show that as a consequence of this $\dinf$ is equal to $\tlambda$.

We define $e_n=[0\hdots 0\,1\,0\hdots 0]^{T}$ where the $1$ is at the $n^{\text{th}}$ position. The nodes reachable on traversing $k$ edges from the initial node $n_0$ are given by the vector $v_k=(\widetilde{A}^T)^k e_{n_0}$, in which each entry of the vector is interpreted as the number of times the corresponding node is reached on traversing $k$ edges.
Since $\tlambda$ is the Perron-Frobenius eigenvalue, and hence the eigenvalue with the largest magnitude, in the limit of $k\to\infty$ the vector $v_k$ asymptotically approaches $(\tlambda)^k v^{PF} \propto v^{PF}$. Hence $\dinf$ is just the average degree calculated using $v^{PF}$ as the set of weights. 

Now, since the degree of node $i$ is given by $\sum_j {\widetilde{A}^T}_{ji}$, we can write 

\begin{align}
\dinf &= \frac{\sum_i d_i v^{PF}_i}{\sum_i v^{PF}_i}, \nonumber \\
       &= \frac{\sum_{i,j} {\widetilde{A}^T}_{ji} v^{PF}_i}{\sum_i v^{PF}_i}, \nonumber\\
       &= \frac{\sum_j \tlambda v^{PF}_j}{\sum_i v^{PF}_i}, \nonumber \\
       &= \tlambda.  \label{eq:firstlevelbard}
\end{align}


\section{Approximation of $\delta r$ via the network coefficients $Q_{\alpha}$} \label{apx:deltar}

Recall that $Q_{\alpha}(n_0)$ is the number of nodes that are reachable from node $n_0$ via exactly $\alpha$ edge-independent paths of length $2$.
We further define $Q_{\alpha}(n_0|\mathcal{V}^0_k)$ to be the number of nodes that are reachable from node $n_0$ via exactly $\alpha$ edge-independent paths of length $2$, \emph{such that the first edge leads to a node in $\mathcal{V}^0_k$}. Assuming that for each $\alpha$ the paths leading to the nodes being counted are distributed uniformly across all paths of length $2$,
\begin{equation}\label{eq:Deltaalpha}
Q_{\alpha}(n_0|\mathcal{V}^0_k) = Q_{\alpha}(n_0) \frac{\binom{D[\mathcal{V}^0_k]}{\alpha}}{\binom{d(n_0) \langle d(n_1) \rangle}{\alpha}}.
\end{equation}
We use these $Q_{\alpha}(n_0|\mathcal{V}^0_k)$ coefficients to estimate the quantity $\delta r$. 

Recall that $r$ edges are activated out of a maximum possible of $D[\mathcal{V}^0_k]$ edges starting from nodes in $\mathcal{V}^0_k$. 
Since each of the $D[\mathcal{V}^0_k]$ edges are equally likely to be chosen, we assume the probability of choosing any one edge is given by $(r/D[\mathcal{V}^0_k])$. 
Now, for any given $\alpha$, let us consider a set of $\alpha$ edges that start from nodes in $\mathcal{V}^0_k$ and end at a common node. 
Out of the $r$ activated edges, if $\beta \geq1 $ edges are chosen to be activated from this set of $\alpha$ edges, then the contribution to $\delta r$ is $\beta - 1$, since only the number of distinct activated nodes are relevant to $|\mathcal{V}^1_r|=r-\delta r$, and each of the $\alpha$ edges lead to the same node. 
For $\beta =0$, the contribution to $\delta r$ is $0$. 
We can find the expected contribution to $\delta r$ by summing over all possible $\beta$'s weighted with their respective probabilities as
\begin{align*}
 \sum_{\beta \geq 1} & \left[ \binom{\alpha}{\beta} \left( \frac{r}{D[\mathcal{V}^0_k]}\right)^{\beta} \left(1- \frac{r}{D[\mathcal{V}^0_k]}\right)^{\alpha - \beta} \right](\beta - 1) \\
=& \alpha \left( \frac{r}{D[\mathcal{V}^0_k]}\right) - 1 +  \left(1- \frac{r}{D[\mathcal{V}^0_k]}\right)^\alpha.
\end{align*}
Since there are $Q_{\alpha}(n_0|\mathcal{V}^0_k)$ such sets of $\alpha$ edges, we can add their contributions together, and sum over all possible $\alpha$ to write $\delta r$ as 
\begin{dmath*}
\delta r = \sum_\alpha Q_{\alpha}(n_0|\mathcal{V}^0_k) \left[ \alpha \left( \frac{r}{D[\mathcal{V}^0_k]}\right) - 1 + \left( 1- \frac{r}{D[\mathcal{V}^0_k]}\right)^\alpha \right],  \\ 
  = \sum_\alpha Q_{\alpha}(n_0) \frac{\binom{D[\mathcal{V}^0_k]}{\alpha}}{\binom{d(n_0) \langle d(n_1)\rangle}{\alpha}} \left[ \alpha \left( \frac{r}{D[\mathcal{V}^0_k]}\right) - 1 + \left( 1- \frac{r}{D[\mathcal{V}^0_k]}\right)^\alpha \right]. 
\end{dmath*}
If we further approximate the binomial expressions in Eq. (\ref{eq:Deltaalpha}) to write 
\begin{equation*}
Q_{\alpha}(n_0|\mathcal{V}^0_k) = Q_{\alpha}(n_0) \times \left[ \frac{D[\mathcal{V}^0_k]}{(d(n_0) \langle d(n_1) \rangle)}\right]^\alpha,
\end{equation*}
we obtain the approximation described in Eq. (\ref{eq:deltar}) of the text. The motivation for this final step of approximation is to allow each term in the summation of the expression for $\delta r$ to be expanded to powers of $D[\mathcal{V}^0_k]$, which we later use in simplifying various terms.

\section{Expressions for sum of powers of $D[\mathcal{V}^0_k]$ over all possible sets $\mathcal{E}^0_k$}\label{apx:sumDSk}

Recall that $D[\mathcal{V}^0_k] = \sum_{v \in \mathcal{V}^0_k} d(v)$, where $\mathcal{V}^0_k$ was a $k$ element subset of the nodes connected to $n_0$. We show that for arbitrary $q \in \mathbb{N}$,
\begin{dmath}
\sum_{\mathcal{E}^0_k} D[\mathcal{V}^0_k]^q = \sum_{\mathcal{V}^0_k} \left( \sum_{v=0}^{q-2} \binom{q}{q-v}\frac{k!}{(k-v-1)!} \langle d(n_1)^{(q-v)}\rangle \langle d(n_1)\rangle^v   \\
                                   +  \frac{k!}{(k-q)!} \langle d(n_1)\rangle^q\right), \label{eq:DSksumapx}\\
                   =\binom{d(n_0)}{k} \left( \sum_{v=0}^{q-2} \binom{q}{q-v}\frac{k!}{(k-v-1)!} \langle d(n_1)^{(q-v)}\rangle \langle d(n_1)\rangle^v  \\
									 +  \frac{k!}{(k-q)!} \langle d(n_1)\rangle^q\right),  
\end{dmath}
 where $\langle d(n_1)^q\rangle$ is to be interpreted as the average value of the $q^{\text{th}}$ power of the degrees of nodes reachable from $n_0$ after traversing $1$ edge, as opposed to $\langle d(n_1)\rangle^q$, which is the $q^{\text{th}}$ power of $\langle d(n_1)\rangle$, which is the average value of the degrees of nodes reachable from $n_0$ after traversing $1$ edge. We present a sketch of the argument for $q=1$ and $q=2$, which may be extended to arbitrary $q$.

Recall that  $\mathcal{E}^0_k$ is the $k$ element set of activated edges that start at node $n_0$. Since we have assumed that the network has no double edges, hence each of these edges ends at a unique node, and we can treat the sum over all possible sets of edges $\mathcal{E}^0_k$ as a sum over all possible $k$ element sets of nodes $\mathcal{V}^0_k$ that are connected to $n_0$ via a single edge. 
For the case of $q=1$, we now have the summation
\begin{equation*}
\sum_{\mathcal{V}^0_k} D[\mathcal{V}^0_k] = \sum_{\mathcal{V}^0_k=\{v_1, v_2 \cdots v_k\}} \left[ d(v_1) + d(v_2) + \cdots + d(v_k) \right]  .
\end{equation*}
There are a total of $\binom{d(n_0)}{k}$ ways of choosing the $k$ element set of vertices connected to $n_0$ out of the maximum possible number of such vertices, $d(n_0)$. Further, since the summation is performed over possible subsets $\mathcal{V}_k$, each vertex connected to $n_0$ appears the same number of times in the overall sum. Hence, we can replace the summand with the expected value of the summand over all possible sets, which is then just $k$ times the average degree of nodes connected to $n_0$. This gives
\begin{align*}
\sum_{\mathcal{V}^0_k} D[\mathcal{V}^0_k] &= \sum_{\mathcal{V}^0_k} k \times \langle d(n_1) \rangle ,\\
                            &= \binom{d(n_0)}{k} k \times \langle d(n_1) \rangle .
\end{align*}

For the case of $q=2$, we have 
\begin{equation*}
\sum_{\mathcal{V}^0_k} (D[\mathcal{V}^0_k])^2 = \sum_{\mathcal{V}^0_k=\{v_1, v_2 \cdots v_k\}} \left[ d(v_1) + d(v_2) + \cdots + d(v_k) \right]^2  .
\end{equation*}
The square of the sum on the right-hand side of the above expression results in terms that are either of the form $d(v_i)^2$, or of the form $d(v_i)d(v_j)$. There are $k$ terms of the first type, with the degree of each node connected to $n_0$ being represented uniformly. Similarly, there are $k(k-1)$ nodes of the second form, which also appear uniformly for all nodes across the summation over all sets $\mathcal{V}_k$. As earlier, we can replace the summand with it's expectation value before evaluating the summand to obtain
\begin{align*}
\sum_{\mathcal{V}^0_k} (D[\mathcal{V}^0_k])^2 &= \sum_{\mathcal{V}^0_k} k \times \langle d(n_1)^2 \rangle + k(k-1) \times \langle d(n_1) \rangle^2,\\
                                &= \binom{d(n_0)}{k} k \times \langle d(n_1)^2 \rangle + k(k-1) \times \langle d(n_1) \rangle^2 .
\end{align*}

We can use a similar reasoning to argue that for an arbitrary $q \in \mathbb{N}$ we have 
\begin{dmath*}
\sum_{\mathcal{V}^0_k} (D[\mathcal{V}^0_k])^q = \sum_{\mathcal{V}^0_k}\left[ k \binom{q}{q} \langle d(n_1)^q \rangle + k(k-1) \binom{q}{q-1} \langle d(n_1)^{(q-1)} \rangle \langle d(n_1) \rangle  \\+ \cdots \\+  {k(k-1)\cdots (k-q+2)\langle d(n_1)^2 \rangle \langle d(n_1) \rangle^{(q-2)}} + k(k-1)\cdots (k-q+1) \langle d(n_1) \rangle^q \right] ,\\
\end{dmath*}
which then reduces to the final expression in Eq. (\ref{eq:DSksumapx}).

\section{Polynomial expressions for $\pcrit$ for different values of $\alpha_{thr}$}\label{apx:cutoffexpressions}

The smallest value of the cut-off that leads to a result distinct from the locally tree-like approximation is $\alpha_{thr}=2$. For this value of $\alpha_{thr}$ we obtain
\begin{dmath}\label{eq:secondlevelpcritupto2}
1 = p^2\left\{\dinf^5 - \overline{Q_2} \left[ \dinf - p \dinf + p \langle d^2\rangle + p^2 \dinf^2 (\dinf-1) \right] \right\}/\dinf^3,
\end{dmath}
where we follow the same notation as discussed earlier at the end of Sec. \ref{sec:pcrit} and write $\pcrit$ as $p$ for simplicity. At a cut-off of $\alpha_{thr}=3$ we obtain the result shown in Eq. (\ref{eq:secondlevelpcritupto3})
\begin{dmath}\label{eq:secondlevelpcritupto3}
1=\left(\frac{p^2}{d^5}\right)\times \left\{ d^7 - d^2 \overline{Q_2} \left[ d - p d + p \langle d^2 \rangle + p^2 d^2(d - 1)\right] + \overline{Q_2} \left[ -3 \langle d^2 \rangle (p-1)^2 + p \langle d^3 \rangle (p-3) -3 p (p-1)^2 d^2 (d -1) + p^3 (p-3) d^3 (d -1) (d -2) + d (1+ p (-3 + p (2+3 \langle d^2 \rangle (p-3) (d-1))))\right]\right\}.
\end{dmath}

For $\alpha_{thr}=6$ we obtain
\begin{widetext}
\begin{dmath}\label{eq:pcritalphac6}
1= \frac{p^2}{d^{11}} \left(d^{13} \\
+p^2 \left\{(p-3) p \overline{Q_3}-\overline{Q_2}\right\} d^{11} \\
+\left\{d^2 \left[\overline{Q_2}-3 (p-3) p \overline{Q_3}\right] p^2+\left\langle d^2\right\rangle  \left[3 (p-3) p \overline{Q_3}-\overline{Q_2}\right] p+d (p-1) \left[\overline{Q_2}-3 (p-1) p \left(\overline{Q_3}-2 (p-1) p \overline{Q_4}\right)\right]\right\} d^8 \\
+\left\{2 d^3 (p-3) \overline{Q_3} p^3+(p-3) \left\langle d^3\right\rangle  \overline{Q_3} p+3 d^2 (p-1)^2 \left[\overline{Q_3}-6 (p-1) p \overline{Q_4}\right] p-3 (p-1)^2 \left\langle d^2\right\rangle  \left[\overline{Q_3}-6 (p-1) p \overline{Q_4}\right]+d \left[p \left(\{4-11 p\} \overline{Q_4}+5 \{p-1\} p \{7 p-2\} \overline{Q_5}\right) (p-1)^2+\left(p \left\{p \left[-3 p \left\langle d^2\right\rangle +9 \left\langle d^2\right\rangle +2\right]-3\right\}+1\right) \overline{Q_3}\right]\right\} d^6 \\
+\{p-1\} \left\{12 (p-1)^2 p^2 \overline{Q_4} d^3+(p-1) p \left[(11 p-4) \overline{Q_4}-15 (p-1) p (7 p-2) \overline{Q_5}\right] d^2+\left[\left(1-6 \{p-1\} p \left\{3 [p-1] \left\langle d^2\right\rangle -1\right\}\right) \overline{Q_4}+5 (p-1) p \left(\{2 [4-5 p] p-1\} \overline{Q_5}+3 \{p-1\} p \{5 p [3 p-2]+1\} \overline{Q_6}\right)\right] d+[p-1] \left[(4-11 p) \left\langle d^2\right\rangle  \overline{Q_4}+6 (p-1) \left\langle d^3\right\rangle  \overline{Q_4}+15 (p-1) p (7 p-2) \left\langle d^2\right\rangle  \overline{Q_5}\right]\right\} d^4 \\
+\{p-1\} \left\{10 (p-1)^2 p^2 (7 p-2) \overline{Q_5} d^3+5 (p-1) p \left[(2 p \{5 p-4\}+1) \overline{Q_5}-9 (p-1) p (5 p \{3 p-2\}+1) \overline{Q_6}\right] d^2+\left[p \left(-15 \{7 p-2\} \left\langle d^2\right\rangle  \{p-1\}^2+12 p \{2 p-3\}+14\right)-1\right] \overline{Q_5} d-(p-1) p \left[p \left(274 p^2-346 p+109\right)-6\right] \overline{Q_6} d+5 (p-1) \left[(2 \{4-5 p\} p-1) \left\langle d^2\right\rangle  \overline{Q_5}+(p-1) (7 p-2) \left\langle d^3\right\rangle  \overline{Q_5}+9 (p-1) p (5 p \{3 p-2\}+1) \left\langle d^2\right\rangle  \overline{Q_6}\right]\right\} d^2 \\
+\{p-1\} \left\{30 (p-1)^2 p^2 [5 p (3 p-2)+1] d^3+(p-1) p \left[p \left(274 p^2-346 p+109\right)-6\right] d^2+\left[1-15 (p-1) p \left(3 \{p-1\} \{5 p (3 p-2)+1\} \left\langle d^2\right\rangle -2 \{1-2 p\}^2\right)\right] d-(p-1) \left[\left(p \left\{274 p^2-346 p+109\right\}-6\right) \left\langle d^2\right\rangle -15 (p-1) (5 p \{3 p-2\}+1) \left\langle d^3\right\rangle \right]\right\} \overline{Q_6} \\
-p \{p [p (\{p-6\} p+15)-20]+15\} \left\{(d-1) d p \left[(d-2) d p \left(\{d-3\} d p \left\{[d-4] d p \left[(d-5) p d^2+15 \left\langle d^2\right\rangle \right]+20 \left\langle d^3\right\rangle \right\}+15 \left\langle d^4\right\rangle \right)+6 \left\langle d^5\right\rangle \right]+\left\langle d^6\right\rangle \right\} \overline{Q_6} \\
+\left\{(d-1) d p \left[(d-2) d p \left(\{d-3\} d p \left\{(d-4) p d^2+10 \left\langle d^2\right\rangle \right\}+10 \left\langle d^3\right\rangle \right)+5 \left\langle d^4\right\rangle \right]+\left\langle d^5\right\rangle \right\} \left\{15 \overline{Q_6} (p-1)^5+d^2 p [p (\{p-5\} p+10)-10] \overline{Q_5}\right\} \\
-\left\{(d-1) d p \left[(d-2) d p \left(\{d-3\} d p+6 \left\langle d^2\right\rangle \right)+4 \left\langle d^3\right\rangle \right]+\left\langle d^4\right\rangle \right\} \left\{p [(p-4) p+6] \overline{Q_4} d^4+10 (p-1)^4 \overline{Q_5} d^2+5 (p-1)^4 (17 p-4) \overline{Q_6}\right\}\right)
\end{dmath}
\end{widetext}

Such expressions can be easily generated for large values of $\alpha_{thr}$ by analytically evaluating the required sums using Eqs. (\ref{eq:secondlevelrecursion}), (\ref{eq:deltar}) and (\ref{eq:DSksumapx}) via computational tools such as \texttt{Wolfram Mathematica(v 10.2, Student Edition)}, which we used to generate the above expressions. We note however that our requirement for a large value of $\alpha_{thr}$ stemmed from our algorithm used for network generation. In practice, one can easily measure $\overline{Q_\alpha}$ for large values of $\alpha$ for a given network and choose $\alpha_{thr}$ to be sufficiently large to capture most nonzero values of $\overline{Q_\alpha}$; we do not expect large values of $\alpha$ to be required for real-world networks.

\end{document}